\documentclass[twocolumn]{aastex63}

\usepackage[T1]{fontenc}
\usepackage{color}
\usepackage{amsmath,amstext}
\usepackage{natbib}
\usepackage[caption=false]{subfig}

\newcommand{\code}[1]{\texttt{#1}}
\newcommand{\mesa}{\code{MESA}}
\newcommand{\MESA}{\mesa}

\newcommand{\sgra}{Sgr~A$^*$}

\newcommand\beq{\begin{equation}}
\newcommand\eeq{\end{equation}}

\newcommand{\Msun}{\ensuremath{{\rm M}_\odot}}
\newcommand{\mstar}{\ensuremath{M_*}}
\newcommand{\rstar}{\ensuremath{R_*}}
\newcommand{\lstar}{\ensuremath{L_*}}
\newcommand{\mso}{\ensuremath{{\rm M}_\odot}}
\newcommand{\msol}{\ensuremath{{\rm M}_\odot}}

\newcommand{\lso}{\ensuremath{{\rm L}_\odot}}
\newcommand{\kms}{{\rm km}\,\second^{-1}}

\newcommand{\second}{\unitstyle{s}}

\newcommand{\mdotbondi}{\ensuremath{\dot{M}_{\rm B}}}
\newcommand{\mdotbondir}{\ensuremath{\dot{M}_{\rm B, \Gamma}}}
\newcommand{\rbondi}{\ensuremath{R_{\rm B}}}
\newcommand{\rhoagn}{\rho_{\rm AGN}}
\newcommand{\tagn}{T_{\rm AGN}}
\newcommand{\tstar}{T_{*}}

\newcommand{\cs}{c_s}
\newcommand{\csa}{c_{s,{\rm AGN}}}
\newcommand{\csagn}{\csa}
\newcommand{\ledd}{\ensuremath{{\rm L}_{\rm Edd}}}
\newcommand{\vesc}{v_{\rm esc}}
\newcommand{\mdot}{\dot{M}}
\newcommand{\mdotse}{\mdot_{\rm Edd}}
\newcommand{\prad}{P_{\mathrm{rad}}}
\newcommand{\pgas}{P_{\mathrm{gas}}}
\newcommand{\hp}{H_{\mathrm{p}}}

\newcommand{\tauacc}{\tau_{\rm B}} 
\newcommand{\taunuc}{\tau_{\rm Nuc}} 
\newcommand{\tauagn}{\tau_{\rm AGN}}

\newcommand{\unitstyle}[1]{\ensuremath{\mathrm{#1}}}

\newcommand{\centi}{\unitstyle{c}}
\newcommand{\meter}{\unitstyle{m}}
\newcommand{\cm}{\centi\meter}

\newcommand{\agns}{AGN stars}

\begin{document}

\title{Stellar Evolution in the Disks of Active Galactic Nuclei}
\shorttitle{Stellar Evolution in AGN Disks}
\shortauthors{Cantiello, Jermyn, \& Lin}

\author[0000-0002-8171-8596]{Matteo Cantiello}
\affiliation{Center for Computational Astrophysics, Flatiron Institute, 162 5th Avenue, New York, NY 10010,
USA}
\affiliation{Department of Astrophysical Sciences, Princeton University, Princeton, NJ 08544, USA}

\author{Adam S. Jermyn}
\affiliation{Center for Computational Astrophysics, Flatiron Institute, 162 5th Avenue, New York, NY 10010,
USA}

\author{Douglas N.C. Lin}
\affiliation{Astronomy and Astrophysics Department, University of California, Santa Cruz, CA 95064, USA}
\affiliation{Institute for Advanced Studies, Tsinghua University, Beijing, 100086, China}

\correspondingauthor{Matteo Cantiello}
\email{mcantiello@flatironinstitute.org}

\begin{abstract}
Active Galactic Nuclei are powered by geometrically-thin accretion disks surrounding a central supermassive black hole. Here we explore the evolution of stars embedded in these extreme astrophysical environments (AGN stars). Because AGN disks are much hotter and denser than most components of the interstellar medium, AGN stars are subject to very different boundary conditions than normal stars. They are also strongly affected by both mass accretion, which can runaway given the vast mass of the disk, and mass loss due to super-Eddington winds. Moreover, chemical mixing plays a critical role in the evolution of these stars by allowing fresh hydrogen accreted from the disk to mix into their cores. We find that, depending on the local AGN density and sound speed and the duration of the AGN phase, AGN stars can rapidly become very massive (M > 100 $\mso$). These stars undergo core-collapse, leave behind compact remnants and contribute to polluting the disk with heavy elements. We show that the evolution of AGN stars can have a profound impact on the evolution of AGN metallicities, as well as the production of gravitational waves sources observed by LIGO-Virgo. We point to our galactic center as a region well-suited to test some of our predictions of this exotic stellar evolutionary channel.
\end{abstract}

\keywords{stars: evolution, stars: massive, quasars: general, Galaxy: center}

\section{Introduction}\label{sec:introduction}
The discovery of quasars \citep{Schmidt1963} led to a fundamental breakthrough in our
perception of the emergence and evolution of galaxies. These powerful sustained cosmic
beacons are the brightest members of a large population of active galactic nuclei \citep[AGNs,][]{Ho:2008}.  
It is widely accepted that they are powered by the release of gravitational
energy as mass falls onto supermassive black holes via accretion disks 
at ferocious rates ${\dot M}_{bh}$~\citep{LyndenBell1969}.

Over the past five decades, observations have accumulated a vast amount of data on AGNs.
In many cases, AGNs outshine their host galaxies over a wide wavelength range from radio
and IR to UV and X-ray \citep{Elvis1994}. Outflowing jets are commonly found to
originate from AGNs.  Depending on the relative prominence of some observed features,
AGNs have been classified into sub categories.  In an attempt to characterize their
common phenomenological traits and exotic diversity, an empirical unified model has been proposed
\citep{Antonucci:1993,Netzer:2015} for the entire AGN population in terms of the line-of-sight
prospect. In this model, AGNs possess central SMBH (with mass $M_{bh}
\sim 10^{6-10} \Msun$) which are surrounded by geometrically-thin accretion disks
(with aspect ratio $h=H/R \lesssim 10^{-2}$  where $R$ and $H$ are the disk's
radius and scale height) as suggested by \citet{LyndenBell1969}.
Further out, at a distance of light weeks to months from the SMBH, these disks are shrouded by geometrically
thicker layers of fast moving  clouds which produce the broad emission lines seen in some AGN spectra.
A well-puffed-up torus (with $h \gtrsim 10^{-1}$) of cold, molecular gas is located between 
$\approx$ 0.1 - 10 pc \citep{Nenkova:2008}.
Since it is optically thick in the direction of the central SMBH, this torus blocks both
the thin accretion disk and
the broad line regions from view if they are seen edge-on. The so-called ``narrow line region''
is found further out at $\gtrsim 10^2$ pc in the host galaxies. Compared to the broad line
region, these narrow lines are emitted by relatively small, low-density gas clouds moving
at much smaller velocities  \citep{Boroson:1992,
Antonucci:1993, Ferrarese:2005, Ho:2008, Alexander:2017}.

In the last two decades, surveys with XMM, SDSS, and Chandra have greatly expanded the
database of AGNs and provided evidence that SMBHs coevolve with their host galaxies
\citep{Fabian2012, Kormendy2013, Heckman2014}.  The redshift-dependent AGN luminosity
($L$) function has been used to construct population synthesis models \citep{Yu2002,
Marconi2004} to infer the energy-dissipation efficiency factor ($\epsilon = L/
{\dot M}_{bh} c^2$), the ratio $\lambda= L/\ledd$ where $\ledd \approx 3.2 \times 10^4 L_\odot (M_{bh}/\Msun)$ is the electron-scattering Eddington limit
(Eq.~\ref{eq:ledd}),
and the duty cycle of AGNs as functions of the SMBHs' mass ($M_{bh}$)
and redshift ($z$).
These analyses show that the most luminous AGN phases generally last $\sim 10^{8-9}$~yr, although shorter AGN lifetimes are certainly possible \citep{King:2015,Schawinski:2015}.
During these phases, $M_{bh}$ grows substantially with $\lambda \sim 0.6$ and $\epsilon
\sim 0.06$ \citep{Shankar2009,Shankar2013, Raimundo2009}.

Detailed spectroscopic modeling
of the broad emission lines \citep{Nagao2006, Xu2018} indicate that in AGNs, the $\alpha$-element
abundance is 1) generally higher than the solar value, 2) an increasing function of
$M_{bh}$, but 3) independent of redshift.  The detection X-ray florescence 6.4 kev
K-shell emission line \citep{Tanaka1995, Yaqoob1996, Nandra1997} suggests that iron abundance
close to the SMBH may also be substantial, though there are not yet any quantitative constraints.
These observed properties can be explained if heavy elements are produced in
accretion disks either near or outside the broad line regions and accreted into the 
super massive black holes at the center \citep{Artymowicz:1993}.
Indeed observations indicate a link between AGNs and ongoing {\it in situ} star bursts
\citep{Alexander2012}.  It has been hypothesized \citep{Ishibashi2012} that rapid star
formation may also lead to an elevated occurrence of supernovae, an upsurge in the dust
production rate, a reduction in $\ledd$ (due to an increase in the dust opacity), the
clearing of outer regions of the disk, and the quenching of AGN activities.

The common existence of SMBH in normal galaxies is inferred from the measured surface-brightness 
and velocity-dispersion distribution \citep{Kormendy2013}. In contrast, the modest occurrence rate
of AGN activities suggests a low duty cycle ($\sim 10^{-2}$). The centers of typical galaxies are 
quiescent and most SMBH's in galactic nuclei are accreting at a slow rate with $\lambda \ll 1$.

For example, the center of the Milky Way hosts a $M_{bh}
= 4 \times 10^6 \Msun$ \citep{Ghez1998, ghez2003b, ghez2008, Genzel1997,
Genzel:2010, gillessen2009, schodel2009, Boehle2016}.  Within 1~pc from it, there are
$\sim 10^7$ mostly mature stars \citep{Do2009, Genzel:2010}.  There are also $\sim 10^2$
massive OB and Wolf-Rayet (WR) stars residing within $\sim$0.05-0.5 pc around the \sgra\ SMBH
\citep{Krabbe:1995,Genzel:2003,Ghez:2003,LevinB:2003,Alexander:2005,Paumard:2006}.
Their estimated age is $\sim 4-6$ Myr \citep{Ghez2003a}.  Since a fraction of these young
and massive stars lie in one or more orbital plane \citep{lockmann2009}, it is often
assumed that they were either formed \citep{Goodman:2003, LevinB:2003} or captured and rejuvenated \citep{Artymowicz:1993,Davies2020} in
a common disk around the central SMBH.  Although there are stringent upper
limits on the presence of gas in the vicinity of the SMBH in the Galactic
Center today \citep{Murchikova2019}, the existence of a disk, once active during or after
the formation of the massive stars, may be inferred
from the ``Fermi Bubble'' above and below the Galactic plane \citep{Su2010}.
Such a disk may have $\lambda$ comparable to that of accretion disks around
similar-mass SMBH in AGNs.  By analogy with metal-rich AGNs, the evolution of 
massive stars may lead to heavy-element enrichment near  \sgra. Super-solar metallicity
has indeed been detected in the spectra of some stars within 0.5~pc from the Galactic
Center \citep{Do:2018}.

The possibility of stars being born inside, or captured by, an AGN disk has been extensively discussed in
the literature \citep[See e.g.][]{Syer:1991,Artymowicz:1993,Collin:1999,Levin:2003,Goodman:2004,
Nayakshin:2005,Collin:2008,Wang:2011,Mapelli:2012,Dittmann:2020}. However, the problem of the evolution of
stars embedded in AGN disks has not been addressed in detail.  One of the assumptions behind
the conventional theory of stellar evolution is that stars evolve in tenuous, low
temperature gas, i.e. the interstellar medium.  This assumption sets the outer boundary conditions
of the stellar evolution problem, allowing the computation of the stellar structure and its
evolution on the long nuclear burning timescale.  Even in the case of binary interactions, the
stellar properties are usually only modified by episodes of mass and angular momentum exchange
with a stellar companion. While potentially decisive for the final outcome \citep{Langer:2012,demink:2014}, these episodes are
usually short-lived compared to the stellar lifetime. The accepted wisdom is that stars spend
the majority of their lives in a cold vacuum.

In comparison with  the interstellar medium (ISM), the temperatures and densities that can be found in a large volume
of AGN disks are extreme (\S~\ref{sec:agndisk}). Therefore stars embedded in an AGN disk should 
be evolved including very different boundary conditions (\S~\ref{sec:agndiskmodel}) than those 
evolving in the ISM. In particular, AGN stars should be evolved accounting for potentially 
large external temperatures, densities, and accretion rates during large fractions of the stellar 
lifetime (\S~\ref{sec:model}).  In this paper we discuss the adjustments required to stellar 
evolution models (\S~\ref{sec:mesa}) to calculate the structure and evolution of stars embedded
in AGN disks. Using typical AGN disk conditions 
we show in \S~\ref{sec:results} the results of applying these new stellar calculations to
evolve models of stars embedded in different parts of an AGN disk. In \S~\ref{sec:signatures} we discuss possible observational signatures of AGN stars evolution. The final section of the paper summarizes and concludes this work.

\section{Stars in AGN Disks}\label{sec:agndisk}
Broadly speaking, there are two ways for a star to end up in such extreme astrophysical environment: {\it in-situ} formation \citep{Collin:1999,Levin:2003,Goodman:2004,Collin:2008,Wang:2011,Mapelli:2012} and capture \citep{Syer:1991,Artymowicz:1993}.

\subsection{{\it In-Situ} Formation}

Extended AGN accretion disks are expected to become self-gravitating and unstable to fragmentation \citep{Paczynski:1978,Kolykhalov:1980,Shlosman:1987,Goodman:2003}. It has been suggested that some
gravitationally unstable AGN disks are likely to produce stars, with theoretical predictions showing
a preference for the formation of massive objects \citep{Levin:2003,Goodman:2004,Dittmann:2020}. If these stars
are able to further accrete after formation, then they can eventually become supermassive \citep{Goodman:2004}.
For this channel it is important to note that, although gravitational instabilities strongly amplify spiral structures
and lead to torque-induced angular momentum transfer, they don't necessarily induce fragmentation and produce young
stars unless the cooling time of the disk is comparable to or shorter than the local dynamical time
scale \citep[][but see also \cite{Hopkins:2013}]{Gammie2001}.

\subsection{Capture}

\begin{figure}[ht!]
\begin{center}
\includegraphics[width=1.0\columnwidth]{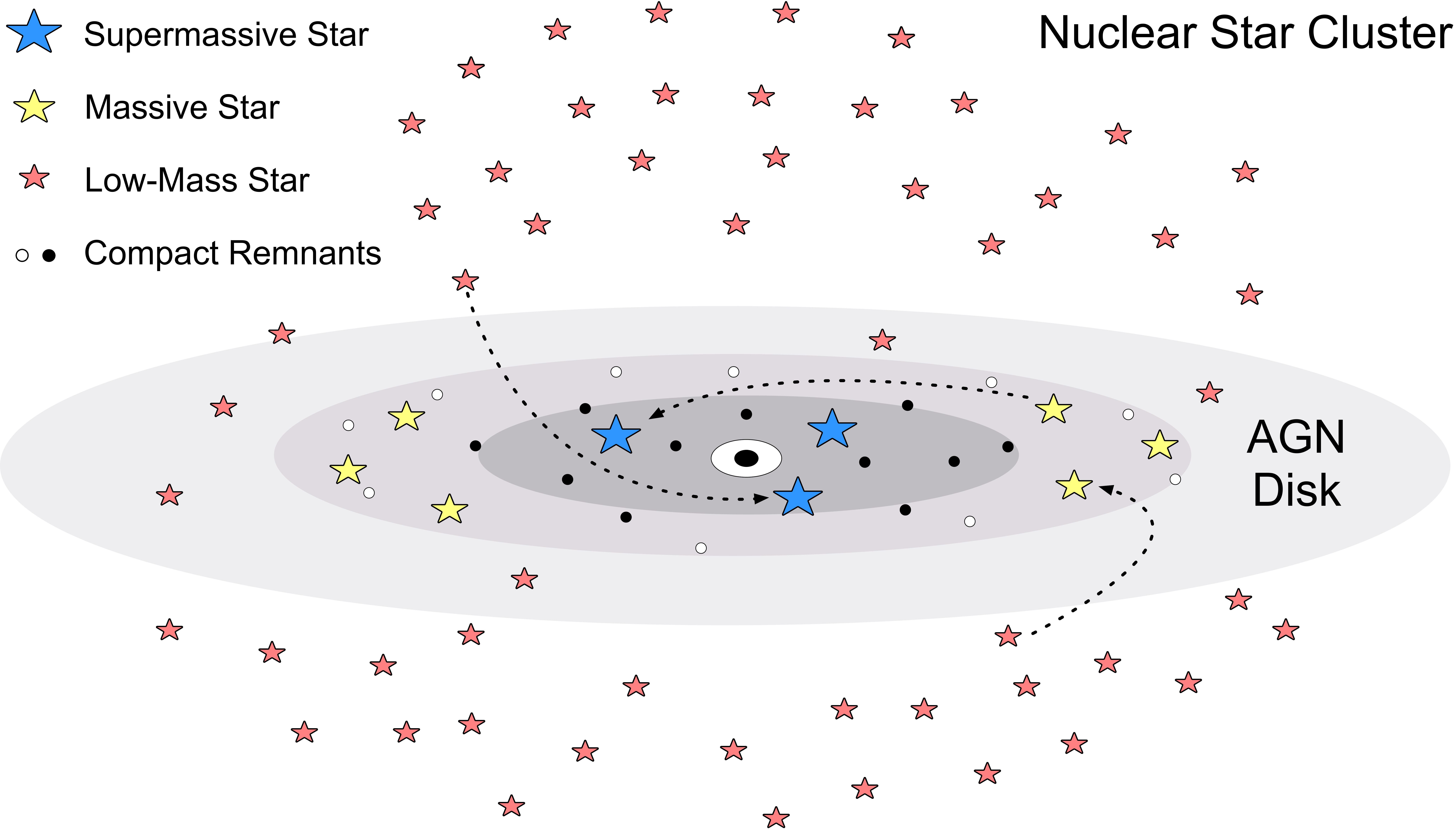}
\caption{\label{fig:agnstars} Schematic representation of stellar evolution in AGN disks. Stars in the inner few parsecs of the disk are either the result of capture from a nuclear star cluster or {\it in-situ} formation. Their evolution is strongly affected by the local AGN conditions, and different regimes of accretion are possible (slow, intermediate and runaway. See \S~\ref{sec:results}). In the inner regions, intermediate and runaway accretion result in a population of massive (M > 8 $\mso$) and supermassive stars (M $\gg$ 100 $\mso$), as well as compact remnants.}
\end{center}
\end{figure}

Stars orbiting the central regions of galaxies can interact with the disk around a SMBH. This interaction results in energy and momentum loss which, over many passages, can bring a star into a circular orbit corotating with the disk \citep{Syer:1991,Artymowicz:1993}. This trapping process relies on hydrodynamical drag as well as the excitation of resonant density waves and bending waves, and can be  an efficient mechanism for $r \lesssim 10$ pc \citep{Artymowicz:1993,Just:2012,Kennedy:2016,MacLeod:2020,Fabj:2020}.
Since  the stellar density is expected to be high in these regions ($n_* \sim 10^6$ pc$^{-3}$) a large number of stars can potentially be trapped during periods of AGN activity (Fig.~\ref{fig:agnstars}).  The exact number depends on the AGN lifetime and the adopted AGN disk model. \citet{Panamarev:2018,Fabj:2020} found that up to a few hundred stars can be captured in a $\sim$Myr timescale. Assuming longer AGN lifetimes (100 Myr), \citet{Artymowicz:1993} predict more than $10^4$ stars could be captured by the AGN.

\subsection{Observational Evidence}

Direct observation of the presence of stars in AGN disks is challenging given the overpowering luminosity of the central galactic
regions, especially during periods of activity \citep{Goodman:2004}.  Nevertheless, {\it in-situ} star formation and/or stellar captures
could leave observational signatures in the stellar populations which are easier to observe in galactic nuclei during quiescent states.
For example, compact stellar clusters are  routinely found in the central $\sim5$~pc of  galaxies, and sometimes they co-exist
with SMBHs \citep[See e.g.][]{Wehner:2006,Neumayer:2020}.  In the center of our Galaxy, the young stars are observed to have a top-heavy present day
mass function \citep[e.g.,][]{NayakshinS:2005,Nayakshin:2006,Paumard:2006,Alexander:2007,Bartko:2010}, which provides some
evidence for stellar capture \citep{Artymowicz:1993}, star formation \citep{Levin:2003,Goodman:2004}, and the 
findings in this paper
(\S~\ref{sec:results}).  Another interesting finding is the rarity of mature red giant branch (RGB) stars
and the overabundance of young, early-type stars near the \sgra\citep{Do:2009,Buchholz:2009,Bartko:2010}.
This correlation supports the notion that capture and rejuvination by accretion may have enhanced the young stellar population close to galactic centers at the expenses of the older population \citep{Davies2020},  although other scenarios are possible \citep{Davies:2005,Zajacek:2020}.

Below we first derive the typical gas conditions in AGN disks (\S~\ref{sec:agndiskmodel}), and then discuss models built to calculate the evolution of stars embedded in these exotic environments (\S~\ref{sec:model}). The results of these novel calculations will give us the ability to explore in \S~\ref{sec:signatures} further observational tests for this scenario.

\section{AGN disk model}\label{sec:agndiskmodel}
The assessment of background density $\rho$ and temperature $T$ requires a model for the AGN accretion disk.
In conventional accretion disk theory \citep{Shakura1973, Pringle1981, King2002}, the 
structure of axisymmetric, geometrically-thin accretion disks is approximated through
the separation of radial ($R$) and vertical (normal to the disk plane, $z$) variables 
under the assumptions: 1) hydrostatic equilibrium in the direction normal to the disk
plane, 2) local thermal equilibrium between the heating due to viscous dissipation
and radiative cooling rates at each radii, and
3) {\it ad hoc} prescriptions of effective viscosity $\nu$.  These models provide estimates for the angular 
momentum transfer rates which can be used to compute the evolution of the surface
density ($\Sigma$). In a steady state, the radial distribution of $\Sigma$, and the
$(R, z)$ distribution of $T$ and $\rho$ can then be determined in terms of the accretion rate
${\dot M}_d$ onto an SMBH's of mass $M_{bh}$. 

In Appendix~\ref{sec:alphadisk} we derive a steady state geometrically-thin viscous disk model. 
We adopt the mean values of $\epsilon (\sim 0.06$) and $\lambda (\sim 0.6$) 
obtained from previous population synthesis models of AGN's \citep{Shankar2009, Shankar2013}. 
These models are relevant for luminous AGNs with signature flat spectral energy distributions 
\citep{Elvis1994, Sirko2003}. We introduce a dispersion factor $f_m$ to include less active AGNs
and  it is equivalent to $\lambda/0.6$.  Based on this model, we obtain scaling laws and estimate 
ranges of mid-plane densities  ($\rhoagn$) and sound speeds  ($\csa$).  
These values provide external conditions for our AGN star models in \S~\ref{sec:mesa}. Bright AGNs 
have relative small ($\sim 10^{-2}$) duty cycle and most SMBH in the center of galaxies do not 
contain such geometrically thin, opaque disk.  Below some threshold ${\dot M}_d$ (equivalently with
$f_m \ll 1$), cooling become inefficient and thermal instability may lead to the transformation of 
thin, cool disks into geometrically-thick, optically-thin ion tori \citep{Ichimaru1977, Rees1982}. 
There is an alternative class of advection-dominated-accretion-flow (ADAF) models which characterize
gas surrounding the SMBH to be both tenuous and hot \citep{Narayan1994}.
In such environments, star formation or star trapping are unlikely to occur.  Moreover, 
accretion onto embedded stars, if any, would proceed at negligible rate.  Therefore,
the analysis presented here is not applicable.  

\subsection{A generic scaling law for marginally self-gravitating disks around AGNs}
\label{sec:scalinglaw}
The gravitational stability of an accretion disk is measured in terms of
\begin{equation}
    Q={\csa \Omega \over \pi G \Sigma} = {h M_{bh} \over {\sqrt 2} \pi \Sigma R^2} 
    = {M_{bh} \over {\sqrt 8} \pi \rhoagn R^3}
    \label{eq:qvalue0}
\end{equation}
where  $h=H/R$ is the aspect ratio, $\Sigma$ is the surface density, and   
$\Omega (= {\sqrt {G M_{bh} / R^3}})$ is the Keplerian angular frequency at radius $R$.
Regions of the disk with $Q \gtrsim 1$ are stable and the midplane density $\rhoagn$ can be
approximated with a steady-state $\alpha$-viscosity prescription (see Appendix~\ref{sec:alphadisk}).
In the gas-pressure dominated region, $\rhoagn \propto r_{\rm pc}^{-33/20}$ where $r_{\rm pc} 
= R/ 1 {\rm pc}$ (Eq.~\ref{eq:rhoagas}).  For a constant opacity, $Q \propto r_{\rm pc}^{-27/30}$ such
the outer regions of the disk are more prone to gravitational instability (GI).  Marginal 
stability is attained with $Q \simeq 1$ (Eqs.~\ref{eq:qvalue0} and \ref{eq:rhoagas}) at 
\begin{equation}
    R_Q \simeq 0.02 \left( {\kappa^3 \alpha^7 \over \mu^{12}} \right)^{2/27} 
    {m_8^{1/27} \over f_m^{8/27}} {\rm pc} .
\end{equation}
where $\mu$, $\kappa$, and $\alpha$ are the molecular weight, opacity, and efficiency of 
turbulent viscosity, $m_8 = M_{bh}/10^8 \Msun$ and $\Msun$ are SMBH's normalized mass and the Sun's mass.
A similar expression can be obtained for the radiation pressure dominated region.  

At radius $R \gtrsim R_Q$, the conventional disk-structure models need to 
be modified to take into account changes in 1) the gravity normal to the disk plane \citep{Paczynski:1978}, 
2) momentum transfer, and 3) possibly thermal and momentum feedback from newly formed or captured stars.
In regions where marginal gravitational stability can be maintained with $Q \sim 1$, the growth 
of non-axisymmetric structure leads to a torque that transports angular momentum with an equivalent 
$\alpha_{\rm GI} \sim 1$ \citep{Lin1987, Kratter2016}. This efficiency is substantially larger than 
that resulting from MHD turbulence induced by the magnetorotational 
instability (MRI, $\alpha_{\rm MRI} \sim 10^{-3}$) \citep{Bai2011} in the inner disk region where self gravity of the disk is negligible 
(with $Q \gg 1$).  We adopt a convenient prescription 
\begin{equation}
    \alpha=\alpha_{\rm MRI} f_Q + \alpha_{\rm GI}(1- f_Q).
\label{eq:alpha}
\end{equation}
We approximate the transition between GI and MRI with a dimensionless factor $f(Q)$ which steeply increases
from 0 to 1 as $Q$ increases beyond $\sim 1$.
 
The above $\alpha-Q$ relationship eliminates another degree of freedom (in addition to the ${\dot M}_{bh}-M_{bh}$ relation in Eq.~\ref{eq:mdotshankar}).  It is a reasonable approximation provided $Q$ is not much less than unity. 
But even with $\alpha \simeq \alpha_{\rm GI} \sim 1$ (in Eqs.~\ref{eq:Tagas}-\ref{eq:pgas} and \ref{eq:Tarad}-\ref{eq:parad}), 
$Q \ll 1$ at $R$ greater than a few $R_Q$.  In such violently unstable disks, global spiral structure rapidly grows 
and induces angular momentum transfer on a dynamical time scale \citep{Papaloizou1991}.  

However, star formation may also be triggered in these disk regions \citep{Gammie2001}
if their cooling time scale 
\begin{equation}
    \tau_{\rm cool} = {\mathcal{E}_{\rm th} \over Q^-} \lesssim {3 \over \Omega} \ \ \ \ \ \  {\rm where}
    \ \ \ \ \ \ \mathcal{E}_{\rm th} \sim \int_{-H} ^H P_{\rm rad} dz 
    \label{eq:starformtime}
\end{equation}
is the column density of thermal energy , $Q^-$ is the thermal energy flux 
out of the disk (Eq. \ref{eq:qminus}), and $P_{\rm rad}$ is the radiation pressure. 
In the gas ($P_{\rm gas}$)/radiation pressure dominated regions (with 
$\beta_{\rm P}  \equiv P_{\rm gas}/ P_{\rm rad}  \gtrless 1$)
$\mathcal{E}_{\rm th} \sim {\Sigma R_g \tagn / \mu}$ and $\sim {2 a \tagn ^4 H / 3}$
respectively. Both limits can be taken into account with 
\begin{equation}
    \mathcal{E}_{\rm tot} \sim \frac{2}{3} (1+\beta_{\rm P}) \, a \tagn^4 H
\end{equation}  
where $\tagn$ is the mid-plane temperature.
The cooling time $\tau_{\rm cool}$, in the opaque (where the optical depth of the disk
$\tau \gg 1$) limit, is
\begin{align}
    \tau_{\rm cool} & \simeq {4 (1+\beta_{\rm P}) H \tau \over 3 c} 
    = {4 {\sqrt 2} (1 + \beta_{\rm P})  c_{\rm s, a} \tau \over 3 c \Omega}
    \label{eq:tcool}
\end{align}
(Eq.~\ref{eq:betap}).
In thermal equilibrium where the local viscous dissipation is balanced by the radiative diffusion,
$\tau_{\rm cool} \sim 1/ \alpha\Omega$ \citep{Pringle1973}.  In this case, the necessary condition for star
formation (i.e. Eq.~\ref{eq:starformtime}) is attainable in the limit of marginal gravitational 
stability ($Q \sim 1$) when $\alpha \sim 1$ (Eq.~\ref{eq:alpha}).

The onset of star formation leads to additional momentum and energy sources such as stellar luminosity,
wind, supernovae, and accretion onto their remnants.  These processes increase $T_{\rm AGN}$, $\csa$ (the mid-plane sound speed), and $H$.
They also increase $Q$ and $\tau_c$ as well as reduce $\alpha$ and therefore reduce the heating rate.  
With a self regulated star formation rate, marginal gravitational stability may be maintained with $Q\sim 1$
outside $R_Q$.   Here we introduce a scaling law to construct a generic model for marginally stable disk regions 
at $R \gtrsim R_Q$.

From the steady state ${\dot M}_{bh}-M_{bh}$ relation (Eq.~\ref{eq:mdotshankar}),
\begin{equation}  
{\dot M}_d = {\dot M}_{bh} = 3 \frac{\alpha h^3}{Q} M_{bh} \Omega  \ \ \ \ \ \ {\rm and}
\end{equation}
\begin{equation}
    h= \left( { {\dot M}_d Q \over 3 \alpha M_{bh} \Omega} \right)^{1/3} = 
    {f_m^{1/3} Q^{1/3} \over \alpha^{1/3}} h_Q, \ \ \ \ \ \ h_Q \equiv {0.025 r_{pc}^{1/2} \over m_8^{1/6}}.
    \label{eq:hq}
\end{equation}
where ${\dot M}_d$ is the mass flux through the disk. 
The corresponding midplane and surface density are  
\begin{equation}
    \rhoagn = {\rho_Q \over Q}, \ \ \ \ \ \
    \rho_Q \equiv {M_{bh} \over {\sqrt 8} \pi R^3} =  8.3 \times 10^{-16} {m_8 \over r_{pc}^{3}} {\rm \ g \, cm^{-3}},
    \label{eq:rhoc}
\end{equation} 
\begin{equation}
\Sigma={f_m ^{1/3} \Sigma_Q \over Q^{2/3} \alpha^{1/3}}, \ \ \ \ \ \ 
\Sigma_Q \equiv  2 \rho_Q h_Q r = {180 m_8^{5/6} \over r_{pc}^{3/2}}\ {\rm \ g \, cm^{-2}}.
\label{eq:rhosigma}
\end{equation}
From the midplane sound speed,
\begin{equation}
    \csa= {h \Omega R \over {\sqrt 2}} \sim {f_m^{1/3} Q^{1/3} \over \alpha^{1/3}} c_{s, Q},
    \ \ \ \ \ \ \ c_{s, Q} \equiv 10^6 m_8 ^{2/3} \ {\rm cm \, s^{-1}},
\label{eq:csa}
\end{equation}
we find the midplane temperature for the gas/radiation pressure dominated regions, 
     \begin{equation}
     \begin{split}
         T_{\rm AGN, gas} =\left({f_m Q \over \alpha} \right)^{2/3} T_{\rm Q, gas},
         \ \ \ \ \ \ T_{\rm Q, gas} \simeq 1.3 \times 10^4 m_8 ^{2/3} {\rm K} \\
    T_{\rm AGN, rad} \sim { f_m ^{1/6} T_{\rm Q, rad}\over \alpha^{1/6} Q^{1/12}},
    \ \ \ \ \ \ 
    T_{\rm Q, rad} \sim 1.4 \times 10^3 { m_8^{5/12} \over r_{pc}^{3/4}} {\rm K} .
    \label{eq:prad}
    \end{split}
\end{equation}

If marginal gravitational stability is maintained outside $R_Q$, the 
outer regions of the disk would have $Q \sim 1$ and $\alpha \sim 1$
with $h \sim h_Q$, $\rhoagn \sim \rho_Q$, $\csa \sim c_{s, Q}$, 
$T_{\rm AGN, gas} = T_{\rm Q, gas}$ and $T_{\rm AGN, rad} = T_{\rm Q, rad}$.
Equations~(\ref{eq:hq}), (\ref{eq:rhoc}), (\ref{eq:csa}),
and (\ref{eq:prad}) provide a range densities and sound speeds for the
AGN disk.  
Note that in Equation~(\ref{eq:csa}), $\csa$ is independent of $R$. 

\subsection{Additional heating mechanisms}

Similar to protostellar disks \citep{Garaud2007}, irradiation on the disk surface can significantly 
increase heating flux ($Q^+$), effective temperature on the disk surface
($T_e$), and  mid-plane ($\tagn$) in the outer regions of the disk.
In addition, embedded stars, formed {\it in situ} or captured by the disk, provide additional heat sources. 
Moreover, their neutron star or black hole byproducts continue to accrete disk gas and radiate
due to the dissipation of gravitational  
energy.   In the present investigation on the evolution of stars in AGN disks,
we consider regions of ongoing star formation or star trapping with $Q \sim 1$ and 
$\alpha \sim \alpha_{\rm GI} \sim 1$.  For the initial exploration of a generic set of 
model parameters, we adopt $\tagn \,(10^{2-3}$ K), $\rhoagn\,(10^{-17} - 10^{-15}$ g cm$^{-3}$), and 
$\csa$ ($3-100~\kms$). Additional models with a more extended 
range of disk properties will be examined in subsequent papers. 
The maintenance of a thermal equilibrium in such disk regions requires 
``auxiliary energy sources'' to the viscous dissipation in conventional disk models.
Some possible contributors include nuclear fusion from embedded stars or accretion power 
from their remnants \citep{Goodman:2003, Thompson2005}.  These sources may also significantly
modify AGN's spectral energy distribution \citep{Sirko2003}.
The construction of a self-consistent stellar feedback model,
including the results in this paper, will be presented elsewhere.

\section{Modelling Stellar Evolution in AGN Disks}\label{sec:model}

As discussed in \S~\ref{sec:agndisk}, theoretical arguments and observations suggest that stars can find themselves inside AGN disks. In this section we discuss the physical ingredients that we implemented in our models to simulate the evolution of such objects, which we call \agns.

\subsection{Accretion}

A star of mass $\mstar$ at rest relative to gas of density $\rhoagn$ and temperature $\tagn$ inside an AGN disk accretes material at a rate
\begin{equation}
\mdotbondi =  \eta \, \pi \rbondi^2 \,\rhoagn\, \csa,
\label{eq:mbondi}
\end{equation}
where $\csa$ is the local sound speed in the disk,
$\rbondi$ is the Bondi radius defined as
\begin{equation}
\rbondi = \frac{2G\mstar}{\csa^2},
\label{eq:rbondi}
\end{equation}
and $\eta$ is an efficiency factor ($\eta \le 1$).

The amount of material that is actually accreted depends on the ability of the inflow to lose angular momentum and on the radiation feedback from the accreting object. 
While the need to shed angular momentum can severely reduce the accretion rate onto compact objects~\citep{Li:2013,Roberts:2017,Inayoshi:2018}, the ratio $\rbondi/\rstar \sim 10^{2...4}$ is much smaller for \agns\ than for BHs and NS ($\sim 10^8$) and so this is less of a barrier to accretion.  We will return to discussing the radiation feedback on accretion in Sec.~\ref{sec:interplay}.

In our models we use Eq.~\ref{eq:mbondi} with an accretion efficiency $\eta = 1$.
However, since  $\mdotbondi \propto \eta \rhoagn/\csa^3$ ($\propto \eta \rhoagn/\tagn^{3/2}$), a different choice of $\eta$ is equivalent to a rescaling of the local AGN conditions. For example, for a fixed $\tagn$, an AGN star model with $\rhoagn = 10^{-16}$ g cm$^{-3}$ and $\eta=1$ is identical to one with $\rhoagn = 10^{-15}$ g cm$^{-3}$ but an accretion efficiency $\eta = 0.1$.
We also assume that the star accretes material with a fixed composition ($\mathrm{X} = 0.72$, $\mathrm{Y} = 0.28$), and that the entropy of the accreted material is the same as that of the stellar surface.
The latter assumption corresponds to an accretion process in which advection is slower than thermal equilibration.
In Appendix~\ref{sec:thermal_adv} we demonstrate that this holds.

\begin{figure}[ht!]
\begin{center}
\includegraphics[width=1.0\columnwidth]{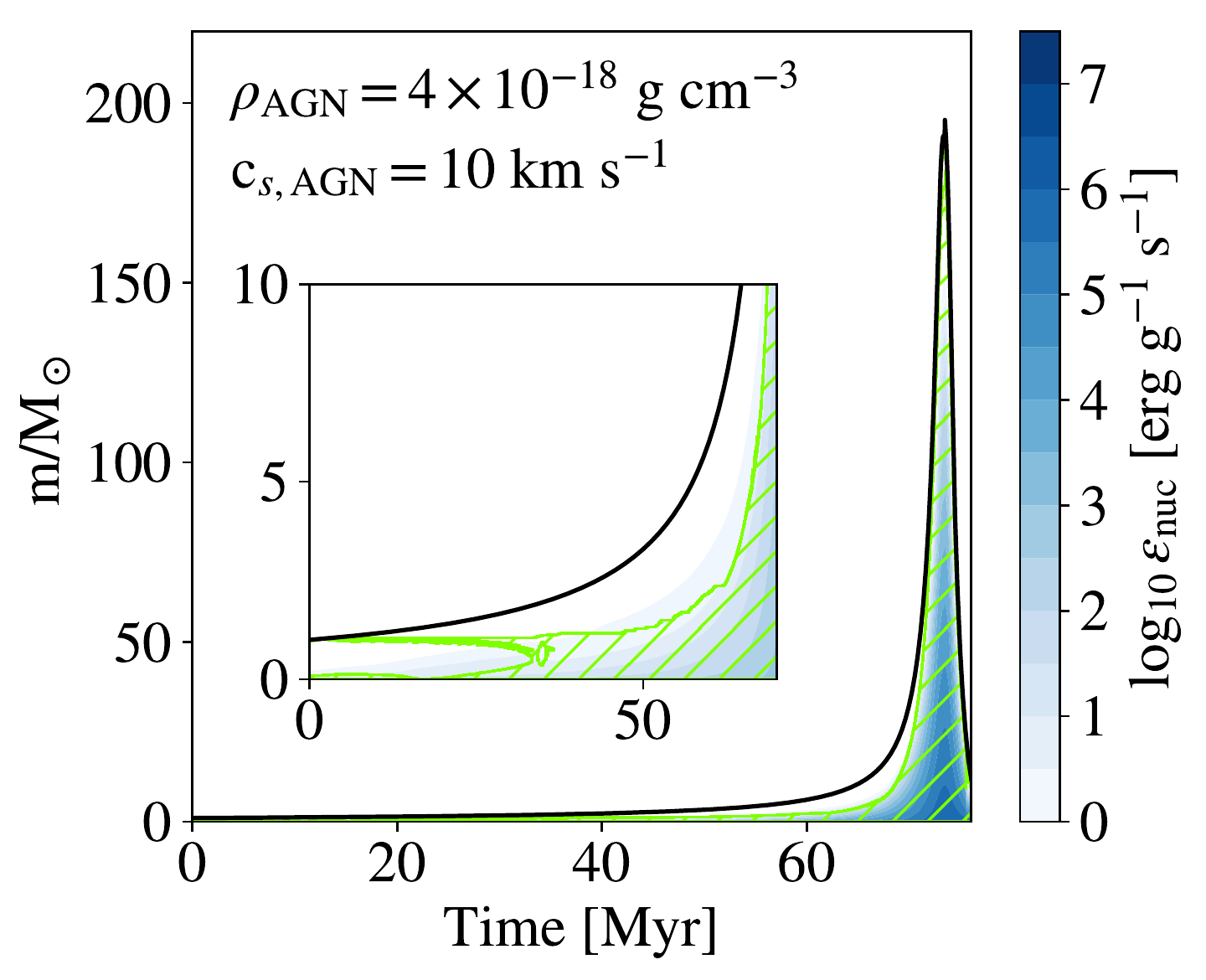}
\caption{\label{fig:kipp4d-18} Evolution of total mass as a function of time for a model of an AGN star at a density of $4\times10^{-18}$ g cm$^{-3}$ and sound speed 10 $\kms$ respectively. Green hatched regions are convective and the blue shading shows the rate of nuclear energy generation. The inset shows a zoom on the first 70 Myr of evolution.}
\end{center}
\end{figure}

\begin{figure*}[ht!]
\begin{center}
\includegraphics[width=0.85\paperwidth]{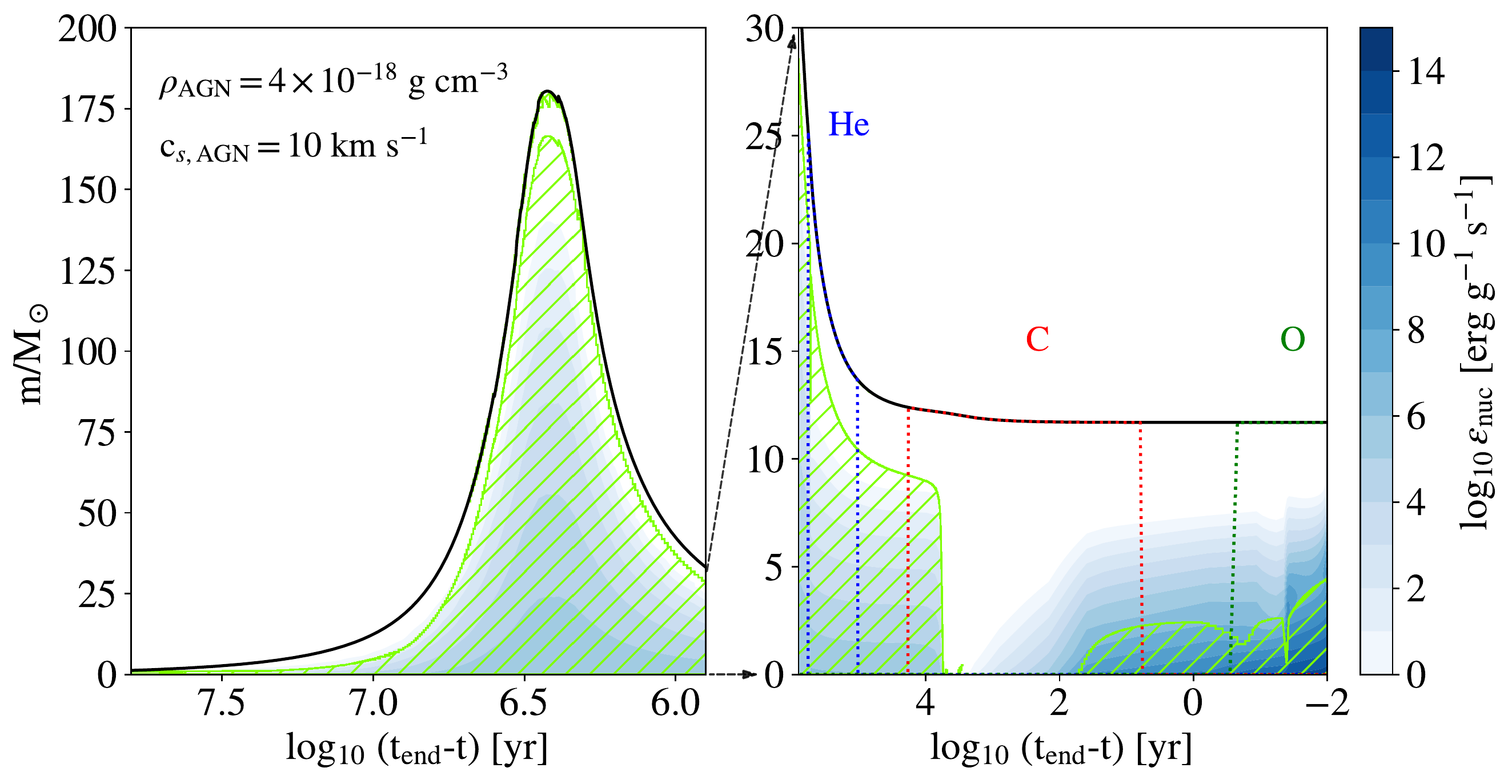}
\caption{\label{fig:kipp_model_4d-18}  Same as Fig.~\ref{fig:kipp4d-18}, but as a function of logarithm of time until the end of calculation (Oxygen burning). The right panel highlights late stages of evolution, showing how the model composition becomes first helium-, then carbon-, and finally oxygen-rich. The calculation reached a core temperature $\log_{10}T_c/$K$\, \simeq 9.5$.}
\end{center}
\end{figure*}

For the range of $\rhoagn$, $\tagn$, and $\csa$ we are considering for AGN disks, ${\dot M}_B$ is comparable
to or larger than that in star-forming dense molecular cores\footnote{These have densities of order $\sim 10^{-20}$ g cm$^{-3}$ and temperatures of order $\sim 10$K.}.
Since massive stars can emerge within $10^6$ yr in the latter, we expect massive stars to rapidly grow in AGNs.

In Fig.~\ref{fig:kipp4d-18} we show the evolution of total mass as function of time for a model of an AGN star at a density of $4\times10^{-18}$ g cm$^{-3}$ and sound speed 10 $\kms$ respectively (corresponding to a radiation-dominated AGN temperature of about 186 K).  Accretion initially dominates and, after about $73\,\mathrm{Myr}$, the star reaches a mass of approximately $200\,\mso$. We will discuss the subsequent evolution of this model after introducing our implementation of mass loss.

\subsection{Mass loss}\label{sec:massloss}

The large accretion rates that can characterize the evolution of \agns\ result in very massive stars with vigorous nuclear burning.
Hence, many \agns\ reach the Eddington luminosity
\begin{equation}\label{eq:ledd}
\ledd \equiv \frac{4\pi G M c}{\kappa} = 3.2\times10^4 \, \left(\frac{\mstar}{\mso}\right) \, \lso,
\end{equation}
where $\kappa$ is the opacity and we used the electron scattering value to calculate the scaling relation on the right-hand side of Eq.~\ref{eq:ledd}. 
When the stellar luminosity exceeds this limit, high mass loss rates are expected  \citep[e.g.][]{Owocki:2012,Smith:2014}. While the details of this process are complex \citep{Maeder:2000,Owocki:2004,Graefener:2008,Graefener:2011,Quataert:2016,Jiang:2018Nature}, similarly to other works \citep[e.g.][]{Paxton:2011}  we assume a super-Eddington outflow at the escape velocity $\vesc=(2\,G\mstar/\rstar)^{1/2}$, with scale set by the excess luminosity $\ledd - \lstar$.
In particular, we take
\begin{equation}\label{eq:medd}
\mdotse = -\frac{\lstar}{\vesc^2}\left[1 + \tanh\left(\frac{\lstar - \ledd}{0.1 \ledd}\right)\right],
\end{equation}
 a phenomenological form where the $\tanh$ term is used to smooth the onset of mass loss and help the calculations converge.  We have chosen to use $\vesc$ at the stellar radius, which in our models is the shock radius (i.e. the point where the accretion stream merges with the nearly-hydrostatic star).
This is the relevant velocity scale because escaping material begins stationary at $\rstar$ and has to reach at least $\vesc$ to escape. We further note that using the full opacity in Eq.~\ref{eq:ledd} leads to a lower value of the Eddington luminosity, which in turns results in larger mass loss rates.

\begin{figure}[ht!]
\begin{center}
\includegraphics[width=1.0\columnwidth]{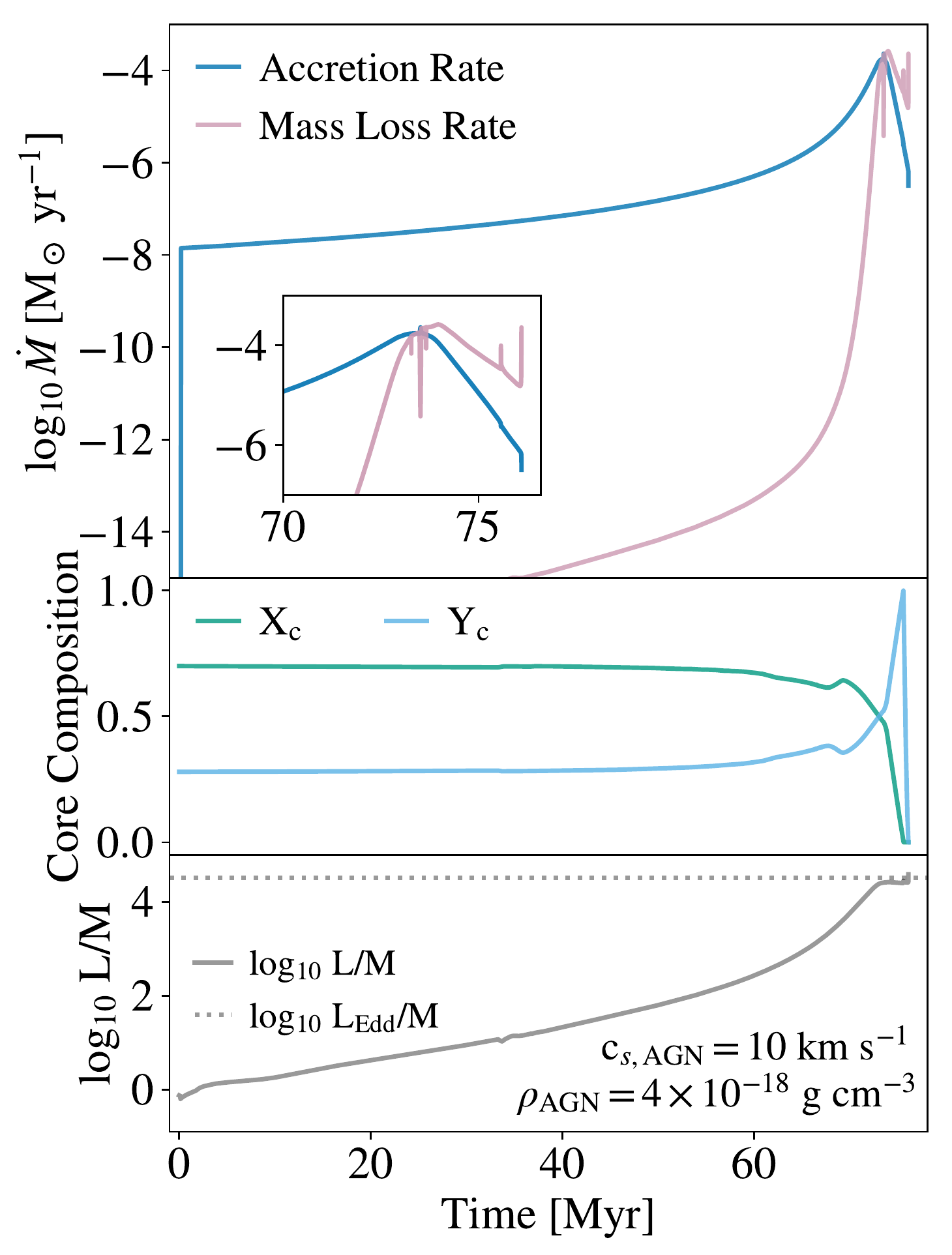}
\caption{\label{fig:accretion4d-18} Time evolution of accretion and mass loss rates in an AGN star model evolved at a density of $4\times10^{-18}$ g cm$^{-3}$ and ambient sound speed 10 $\kms$ (top panel). The core composition is shown in the middle panel (X$_{\rm c}$ and Y$_{\rm c}$ are the core hydrogen and helium mass fractions, respectively). The ratio of stellar luminosity to mass (L/M) in solar units is shown in the bottom panel and compared to the Eddington value.}
\end{center}
\end{figure}

Because of mass accretion and helium enrichment, our model in   Fig.~\ref{fig:kipp4d-18} approaches the Eddington luminosity after $\approx 70$Myr , which drives a super-Eddington stellar wind that eventually dominates over accretion and causes a decrease of its total mass (Fig.~\ref{fig:kipp4d-18}, \ref{fig:kipp_model_4d-18} and \ref{fig:accretion4d-18}). At the end of its evolution the model has a mass of $\approx 10\mso$ (Fig.~\ref{fig:kipp_model_4d-18}), and is expected to undergo core-collapse and leave behind a compact remnant.

As we show in \S~\ref{sec:results}, our models of massive and very massive \agns\ evolve with $\lstar \simeq \ledd$. As such we expect their mass loss to be completely dominated by  super-Eddington, continuum-driven winds, and therefore we do not include any wind mass loss caused by line-driving \citep{Lamers:1999,Vink:2001,Vink:2005,Smith:2014}.

\subsection{Interplay Between Accretion and Mass Loss}\label{sec:interplay}

When \agns\ exceed their Eddington luminosities we expect a complex interplay between accretion and mass loss.
In particular, we expect that turbulent eddies break spherical symmetry, allowing the system to form separated channels of inflows and radiation-dominated outflows~\footnote{In the presence of rotation, such symmetry breaking is likely promoted by the fact that rotating stars have hot poles and cool equators.
In that case, one might expect polar outflows removing mass from the star, while reduced accretion might still occur closer to  equatorial regions.}.
The star should then be both accreting \emph{and} losing mass at the same time.

In our models, we include the effects of accretion and mass loss as follows: for $\lstar < \ledd$ we just add material to the star according to Eq.~\ref{eq:mbondi}. For  $\lstar \ge \ledd$ we first calculate a reduced accretion rate
\begin{equation}
\mdotbondir = \mdotbondi \left(1-\tanh{|\lstar/\ledd|}\right),
\end{equation}
which accounts for the decreased solid angle available to accretion as the star drives a super-Eddington outflow.
This form is purely phenomenological, and was chosen to be smooth and satisfy the constraints of giving the unmodified Bondi accretion rate when $\lstar \ll \ledd$ and zero accretion when $\lstar \gg \ledd$.

\begin{figure}[ht!]
\begin{center}
\includegraphics[width=1.0\columnwidth]{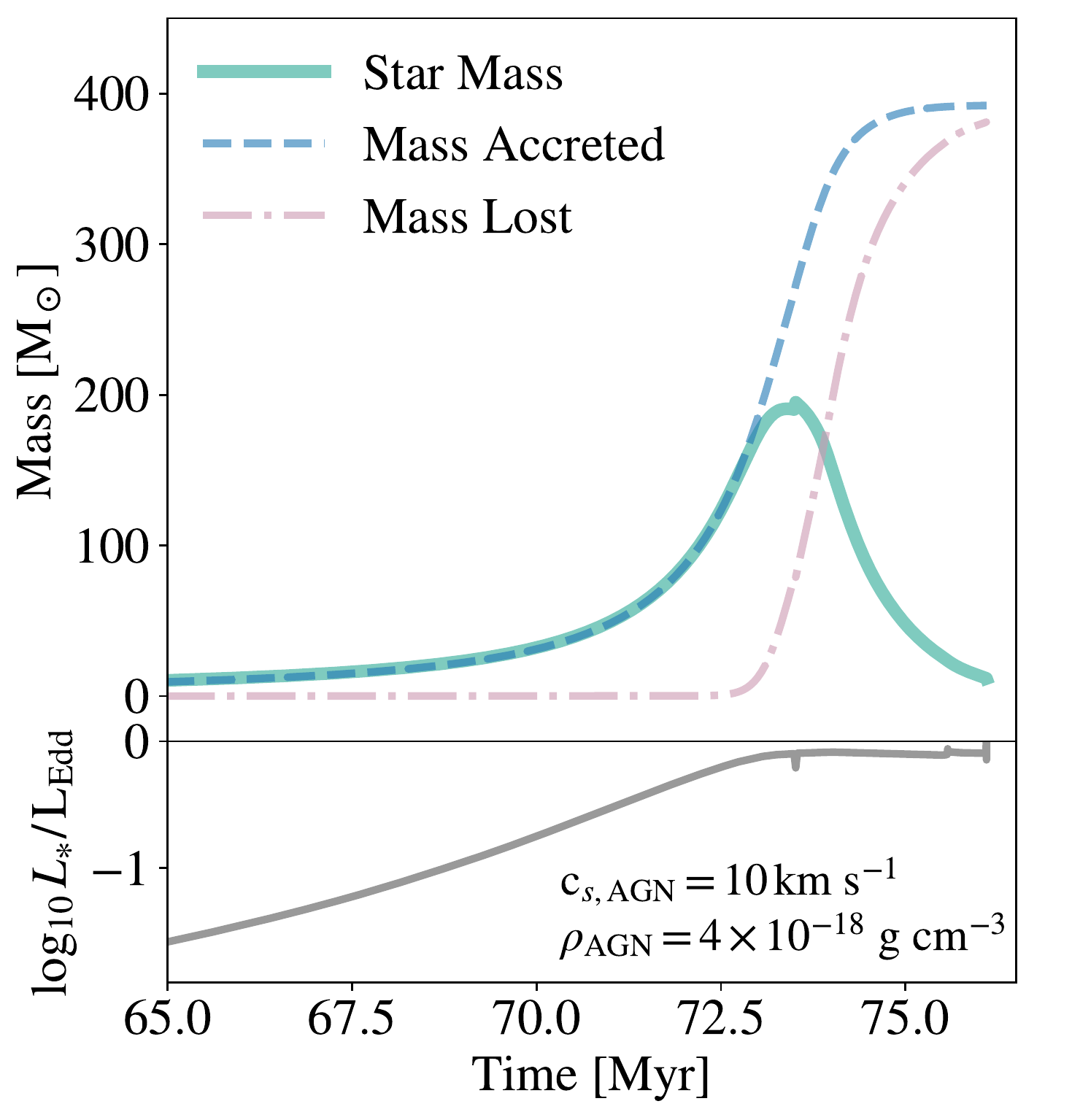}
\caption{\label{fig:gain_loss_4d-18} Mass budget for an AGN star model evolved at a density of $4\times10^{-18}$ g cm$^{-3}$ and a temperature of about $186\,\mathrm{K}$ (AGN sound speed of 10 $\kms$). For the last $\approx$ 12 Myr of calculation, we show the evolution of the model's total mass (green curve), cumulative accreted mass (blue dashed line) and cumulative mass lost (purple dash-dotted line). In the lower panel we also show the time evolution of the Eddington ratio.   
}
\end{center}
\end{figure}

Ordinarily \mesa\ takes a time-step by first changing the mass of a model and then calculating its time evolution through a time-step $dt$ at fixed mass.
To incorporate the effect of simultaneous mass loss and accretion on the composition of our models we modify this procedure so that in each step we first remove an amount of material
\begin{equation}
	\Delta M_{\mathrm{loss}} = \mdotse\,dt,
\end{equation}
and subsequently add an amount of material
\begin{equation}
	\Delta M_{\mathrm{gain}} = \mdotbondir\,dt.
\end{equation}
 The ordering of these allows the \agns\ to release nuclear-processed material back to the AGN \emph{even when} it is net accreting ($\mdotbondir > \mdotse$), see e.g. Fig.~\ref{fig:gain_loss_4d-18}.

This procedure is valid so long as the surface composition of the star is uniform down to a mass-depth $\Delta M_{\rm loss}$.
As we argue in \S~\ref{sec:mixing}, whenever these models undergo super-Eddington mass loss they should also be well-mixed, so this is not a concern for our calculations.

The qualitative evolution of \agns\ close to the Eddington limit does not depend too much on the details of how material is added and removed so long as three qualitative features hold:
\begin{enumerate}
\item \agns\ are able to release nuclear-processed material back to the AGN.
\item \agns\ reaching the Eddington limit lose mass at an increasing rate as $\lstar/\ledd$ increases.
\item Accretion onto \agns\ slows as $\lstar/\ledd$ increases.
\end{enumerate}

\subsection{Internal Mixing}
\label{sec:mixing}

As \agns\ accrete material and become more massive we expect them to become increasingly mixed.
This occurs for several reasons.
First, massive stars have large convective cores which grow as the star becomes more massive.
We shall see that \agns\ also form such cores, which are well-mixed by convection.
Secondly, \agns\ become increasingly radiation-dominated as they become more massive.
This occurs in normal massive stars as well, but because of the unusual surface boundary conditions of stars embedded in AGNs  (\S~\ref{sec:surface_boundary}), the entire star often evolves to a state where $\prad \approx \pgas$.
This means that the star is nearly a $\gamma=4/3$ polytrope, which radically reduces the threshold for any instability to develop, such that even radiative regions can be extensively affected by mixing processes~\citep[see, e.g.][]{Jiang:2015,Jiang:2018Nature}.

Moreover, the interplay of accretion and mass loss is likely to drive strong circulations in the outer radiative part of the star. For example, if mass accretion occurs preferentially in the equatorial regions while mass is lost primarily in the form of polar winds, a strong meridional flow will be driven to restore pressure balance.
This is the case even in the absence of rotational mixing, though rotational mixing is also likely to be important when \agns\ accrete material with significant angular momentum. We will study the impact of rotation on \agns\ evolution in future works.

We model the effect of mixing by adding a compositional diffusivity $D$ that increases with stellar luminosity as
\begin{equation}\label{eq:mixing}
D = \hp \left(\frac{F}{\rho}\right)^{1/3} \, \tanh{\left(\frac{\lstar}{\ledd}\right)^\xi},
\end{equation}
where $F$ is the heat flux and
\begin{align}
	\hp \equiv \frac{P}{\rho g}
\end{align}
is the local pressure scale height.
The form of this additional diffusivity is set to be of order the convective diffusivity \emph{were the region efficiently convectively unstable}. We chose a large value of the exponent ($\xi$ = 7) so that stars become well-mixed only when they get very close to the Eddington limit. With this prescription, \agns\ that accrete large amounts of mass and reach their Eddington luminosity do not build substantial compositional gradients while burning hydrogen in their cores, and evolve quasi-chemically homogeneously \citep{Maeder:1987,Yoon:2005}. On the other hand, \agns\ that only accrete a few solar masses of material evolve mostly as canonical stars. We note that our results depend very weakly on the specific choice of $\xi > 1$.

\begin{figure}[ht!]
\begin{center}
\includegraphics[width=1.0\columnwidth]{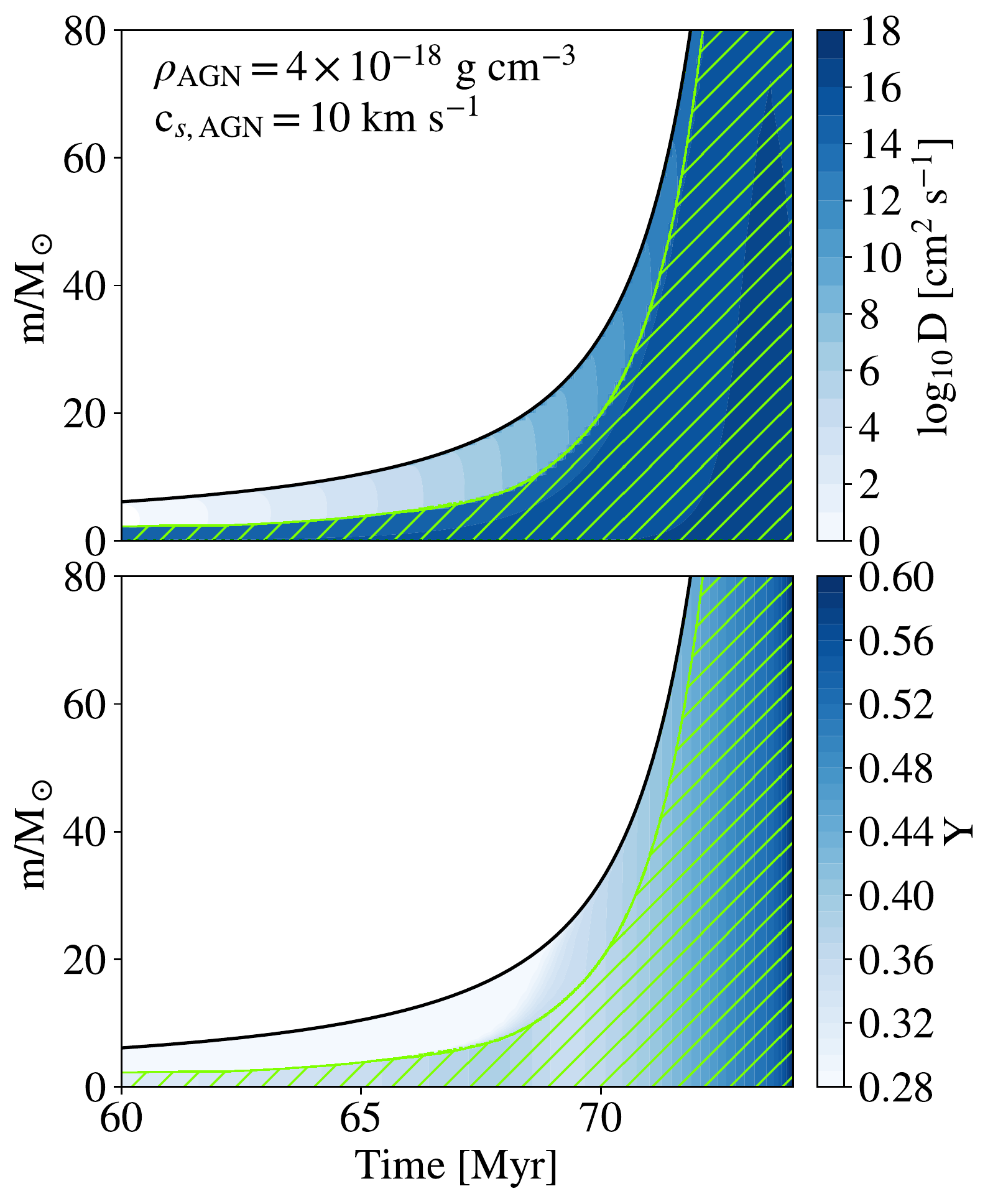}
\caption{\label{fig:kippstack8d-17} Zoom in on the evolution of an AGN star model at a density and ambient sound speed of $4\times10^{-18}$ g cm$^{-3}$ and 10 $\kms$ respectively. The Kippenhahn plots show the increasing role of internal mixing as the stellar mass increases. In the upper panel the blue shading shows the magnitude of the compositional diffusivity D (in cm$^2$ s$^{-1}$), which is always large in convective regions (green-hatched) and increases in radiative regions as the star approaches the Eddington limit (see Eq.~\ref{eq:mixing}). In the lower panel the shading indicates the helium mass fraction (Y).}
\end{center}
\end{figure}

We show in Fig.~\ref{fig:kippstack8d-17} the result of this mixing implementation for our AGN star model with a density and ambient sound speed of $4\times10^{-18}$ g cm$^{-3}$ and 10 $\kms$, respectively. The plots show the increasing role of internal mixing as the stellar mass increases, with an increasing value of the compositional diffusivity leading to mixing of He-rich material outside of the stellar convective core and into the radiative stellar envelope starting at an age of $\approx 67\,\mathrm{Myr}$. This mixing also leads to a partial rejuvination of the star, with H-rich material  supplied to the stellar core (see bumps in the core composition visible at $\approx 67\,\mathrm{Myr}$ in the middle panel of Fig.~\ref{fig:accretion4d-18}). The increased stellar mass and helium abundance in the envelope are responsible for the model reaching the Eddington luminosity. Later in the evolution mass loss is also responsible for revealing to the surface nuclear-processed material, which in turns can have an impact on mass loss itself (by increasing the ratio of stellar luminosity to mass).  The two spikes in mass loss rate visible in the inset of Fig.~\ref{fig:accretion4d-18} (top panel) correspond to the stellar surface abundances becoming helium and oxygen dominated, respectively (see right panel in Fig.~\ref{fig:kipp_model_4d-18}.)

\subsection{Surface Boundary Conditions}
\label{sec:surface_boundary}

Accretion often involves a shock at which material slows from super-sonic to sub-sonic.
To avoid modelling the accretion shock in \MESA, we place the outer boundary of the \MESA\ model just inside the shock.
To do this we must specify the surface pressure and temperature of the model as functions of $\lstar$, $\mstar$, $\rstar$, and the AGN properties.

We derive the structure of the accretion stream, and hence of the surface of the stellar model, by making the following assumptions and approximations:
\begin{enumerate}
	\item The stream is spherically symmetric.
	\item The stream is in steady state.
	\item The stream is not pressure supported.
	\item The stream primarily transports heat via radiative diffusion.
	\item The luminosity is constant in the stream.
	\item The mass of the stream is small compared with $\mstar$.
	\item The opacity of the stream is constant in space, and equal to the electron scattering value.
\end{enumerate}
In Appendix~\ref{sec:assumptions} we verify that the resulting solution either obeys these or is not significantly altered if they fail to hold.

Assumptions (3) and (7) are the most suspect.
Radiation pressure may be significant when $\lstar \approx \ledd$, but we think that even in such cases the answer is unlikely to be very different from what we derive below.
Likewise the opacity almost certainly varies substantially throughout the accretion stream, and our neglect of this variation could introduce systematic errors into our calculations.
Because the opacity only enters in setting a small part of the shock pressure and the total temperature difference between the AGN and the stellar photosphere, our hope is that using a suitable average value is not a bad approximation, but we have no proof to that effect.

\subsubsection{Pre-Shock Properties}

With our assumptions, the inviscid Navier-Stokes equation reads
\begin{align}
	\frac{\partial}{\partial r}\left(\frac{1}{2} v^2\right) + \frac{G \mstar}{r^2} = 0,
	\label{eq:NS}
\end{align}
and mass continuity becomes
\begin{align}
	\dot{M} = 4\pi r^2 \rho v = \mathrm{const.}.
	\label{eq:cont}
\end{align}
Assuming the material is stationary at infinity, we find
\begin{align}
	v \approx \sqrt{\frac{G\mstar}{r}}.
	\label{eq:vscale}
\end{align}
Inserting this into equation~\eqref{eq:cont} we obtain
\begin{align}
	\rho = \frac{\dot{M}}{4\pi r^2}\sqrt{\frac{r}{G\mstar}}.
	\label{eq:rho_scale}
\end{align}

The equation of radiative thermal equilibrium is
\begin{align}
	\frac{dT}{dr} = -\frac{3\kappa \rho L}{64 \pi r^2 \sigma T^3}.
	\label{eq:dTdr}
\end{align}
With fixed $\kappa$ and $L$, and using $\rho \propto r^{-3/2}$, we find that at high optical depth
\begin{align}
	T \propto r^{-5/8}.
	\label{eq:tscale}
\end{align}

\subsubsection{Post-shock Properties}

Because thermal diffusion is much faster than thermal advection over scales of $\rstar$ (see Appendix~\ref{sec:assumptions}) there is no significant temperature jump at the shock.
Rather, the increased entropy is turned promptly into luminosity, of order
\begin{align}
	L_\mathrm{shock} \sim \dot{M} v^2,
	\label{eq:Lshock}
\end{align}
which is added to $\lstar$ to set the total luminosity in the accretion stream.

Because there is no temperature jump, the surface temperature of the star is just that of the base of the stream.
To smoothly connect the limit of an optically thick accretion stream to an optically thin one, we let
\begin{align}
	T_{\rm eff} \equiv \left(\tagn^4 + \frac{L}{4\pi R_{\rm ph}^2 \sigma}\right)^{1/4}
\end{align}
and
\begin{align}
    \tstar = T_{\rm eff} \left(\frac{\rstar}{R_{\rm ph}}\right)^{-5/8},
    \label{eq:Teff}
\end{align}
where $T_{\rm eff}$ is the temperature at the photosphere, $\tstar$ is the temperature of the surface cell, and $R_{\rm ph}$ is the photosphere radius.
To calculate this note that the optical depth between $\rbondi$ and $r$ in the stream is
\begin{align}
	\tau = \int_{\rbondi}^{r} \kappa \rho dr.
\end{align}
Using equation~\eqref{eq:rho_scale} and taking the opacity to be fixed, we find far inside $\rbondi$,
\begin{align}
	\tau \approx \kappa \rhoagn \rbondi \sqrt{\frac{\rbondi}{r}}.
	\label{eq:tauscale}
\end{align}
Because we include the effects of incident radiation from the AGN in equation~\eqref{eq:Teff}, the photosphere occurs when the optical depth between $\rbondi$ and $R_{\rm ph}$ is $\tau \sim 1$, giving
\begin{align}
    R_{\rm ph} \sim \min\left(\rbondi, \rstar+ \kappa^2 \rhoagn^2 \rbondi^3\right),
    \label{eq:rph}
\end{align}
where we have added $\rstar$ so that when the stream is optically thin, equation~\eqref{eq:rph} gives $R_{\rm ph} = \rstar$ and this reduces to the usual Eddington atmospheric condition that places the photosphere at the edge of the stellar model.
Note that we include the term $\tagn^4$ to account for radiation incident from the AGN, along similar lines to the boundary condition of~\citet{1989MNRAS.238..427T}.

\begin{figure}[ht!]
\begin{center}
\includegraphics[width=1.0\columnwidth]{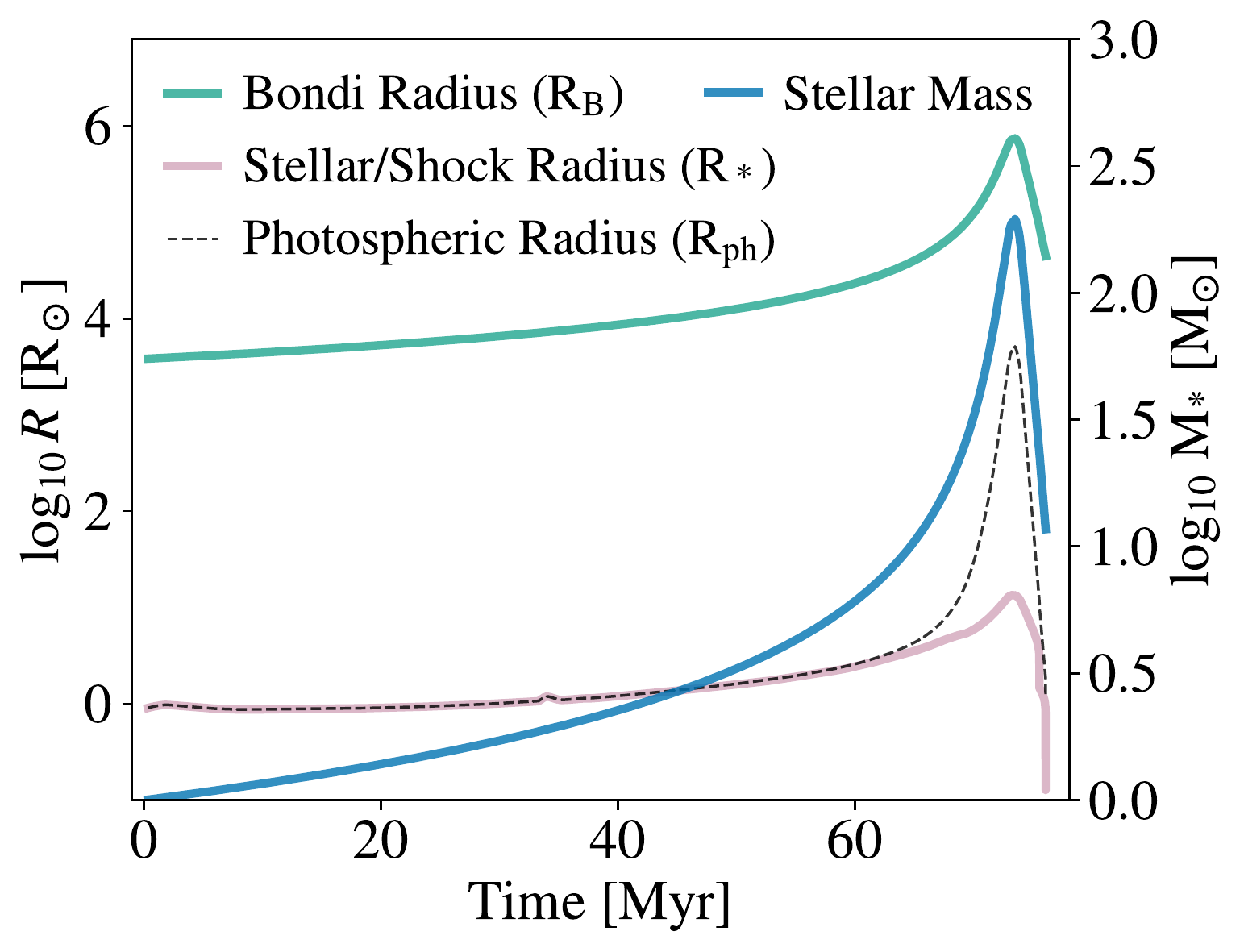}
\caption{\label{fig:radii4d-18} Time evolution of relevant radii for the problem of a star evolving in an AGN disk. The largest radius is always the Bondi radius (Eq.~\ref{eq:rbondi}). 
The outer boundary of the \MESA\ calculation is placed at the shock radius, where the infalling material slows from super-sonic to sub-sonic. The photospheric radius (Eq.~\ref{eq:rph}) corresponds to the shock radius for small accretion rates, but can grow up to the Bondi radius for large accretion rates. Values of these radii are showed, together with total stellar mass, for an AGN star model at a density and ambient sound speed of $4\times10^{-18}$ g cm$^{-3}$ and 10 $\kms$ respectively.}
\end{center}
\end{figure}

To summarize, the stellar radius $\rstar$ corresponds to the outer grid point of our \mesa\ calculation and the location of the accretion shock. The corresponding temperature is $\tstar$. The photospheric radius $R_{\rm ph}$ corresponds to the radius of the star as seen from an observer located at $\rbondi$. The temperature at $R_{\rm ph}$ is $T_{\rm eff}$. We show the evolution of the relevant radii of \agns\ for one of our models in Fig.~\ref{fig:radii4d-18}. Note that, depending on the properties of the surrounding AGN disk, $R_{\rm ph}$ and $T_{\rm eff}$ might not be stellar observables. Finally, for $\mdotbondi \rightarrow 0$,  $R_{\rm ph}\rightarrow \rstar$ and  $T_{\rm eff} \rightarrow \tstar$.

The pressure just after the shock is given by the sum of gas and radiation pressure.
This is related to the pressure just before the shock by the ram pressure, so we write
\begin{align}
	P_{\mathrm{after}} = P_{\mathrm{ram}} + P_{\mathrm{before}}.
	\label{eq:psurf0}
\end{align}
As we mentioned, the thermal diffusivity is high, so the temperature jump across the shock is very small.
As a result the radiation pressure is nearly the same on either side of the shock, so equation~\eqref{eq:psurf0} relates the gas pressure before and after.

The ram pressure due to inflows is
\begin{align}
	P_{\mathrm{ram,inflow}} \approx \rho v^2 \approx \frac{G \mstar \rho}{\rstar}.
\end{align}
When the AGN star loses mass there is a similar term accounting for the force required to launch the outflow:
\begin{align}
	P_{\mathrm{ram,outflow}} = \rho_{\mathrm{outflow}} \vesc^2 = \frac{\mdotse \vesc}{4\pi \rstar^2}.
\end{align}
To ensure that this boundary condition smoothly reduces to the usual Eddington atmospheric condition we also add a contribution $(1/3)\,a \tstar^4 + g/\kappa_{\mathrm{surface}}$, so in full
\begin{align}
	P_{\mathrm{surf}} = \frac{1}{3} a\tstar^4 + \frac{G \mstar \rho}{\rstar} + \frac{\mdotse\vesc}{4\pi \rstar^2} + \frac{g}{\kappa_{\mathrm{surface}}},
\end{align}
where $P_{\rm surf}$ is the pressure in the outermost cell of the \mesa\ model.
Note that we neglect the gas and radiation pressure in the accretion stream and just use the ram pressures because we have assumed that the stream is not pressure supported and hence that the total ram pressure exceeds gas and radiation pressures.

.\section{Numerical Implementation}\label{sec:mesa}

We implemented the physics described in \S~\ref{sec:model} using the Modules for Experiments in Stellar Astrophysics
\citep[MESA][]{Paxton:2011, Paxton:2013, Paxton:2015, Paxton:2018, Paxton:2019} software instrument. Details of the numerical implementation, together with the microphysics inputs to this software instrument are given in Appendix~\ref{sec:software}. 

In the next section we show calculations of \agns\ evolution with ambient conditions $\csagn$ and $\rhoagn$ as derived in \S~\ref{sec:agndiskmodel}. 
We also verify that in our \MESA\ implementation the evolution of \agns\ reduces to classic stellar evolution for  $(\tagn$,$\rhoagn)\to 0$. 

The most important physics affecting the evolution of \agns\ is accretion. The initial accretion rate depends on the AGN conditions $\csagn$  and $\rhoagn$, and the initial mass of the star.  For numerical reasons we assume an initial time of $10^4$ yr during which the accretion rate slowly increases to the Bondi value and the canonical surface boundary conditions are blended with the modified boundary conditions described in \ref{sec:surface_boundary}.
All of our models begin with an initial mass of $1 \mso$ and an initial metallicity of $Z=0.02$. 
We find that different choices of initial mass $M > 1 \mso$  do not change our results significantly.

For these models we also assume that \agns\ remain at constant $\csagn$  and $\rhoagn$ throughout their lives, and that the composition of accreted material is constant with $X=0.72, Y=0.28$ and $Z=0$. Different composition of the accreted material might change results, and we will explore this effect in future work. 
We do not account for migration through the disk or stellar feedback beyond the super-Eddington winds accounted in our models (\S~\ref{sec:massloss}).

\section{Results}\label{sec:results}
There are three important timescales regulating the evolution of \agns: the accretion timescale $\tauacc = M_*/\mdotbondi$, the nuclear timescale $\taunuc \approx 0.007\, M_*c^2 / L_*$, and the AGN lifetime  $\tauagn$.  The evolution of \agns\ is dictated by the hierarchy of these timescales. 

\subsection{Slow Accretion}

When $\tauacc > \taunuc$ or $\tauacc > \tauagn$, \agns\ accrete small or negligible amounts of mass. 
Their evolution is not substantially altered, but depending on the AGN disk conditions a population of low- and intermediate-mass stars with AGN-like photospheric chemistry could be formed.
This population of long-lived \agns\ is interesting because could provide observational tests of the theory as well as a probe of former AGN conditions (see \S~\ref{sec:signatures}).

\subsection{Intermediate Accretion}\label{sec:intermediate}

\begin{figure*}[htp!]
  \centering
  \subfloat{\includegraphics[width=1\columnwidth]{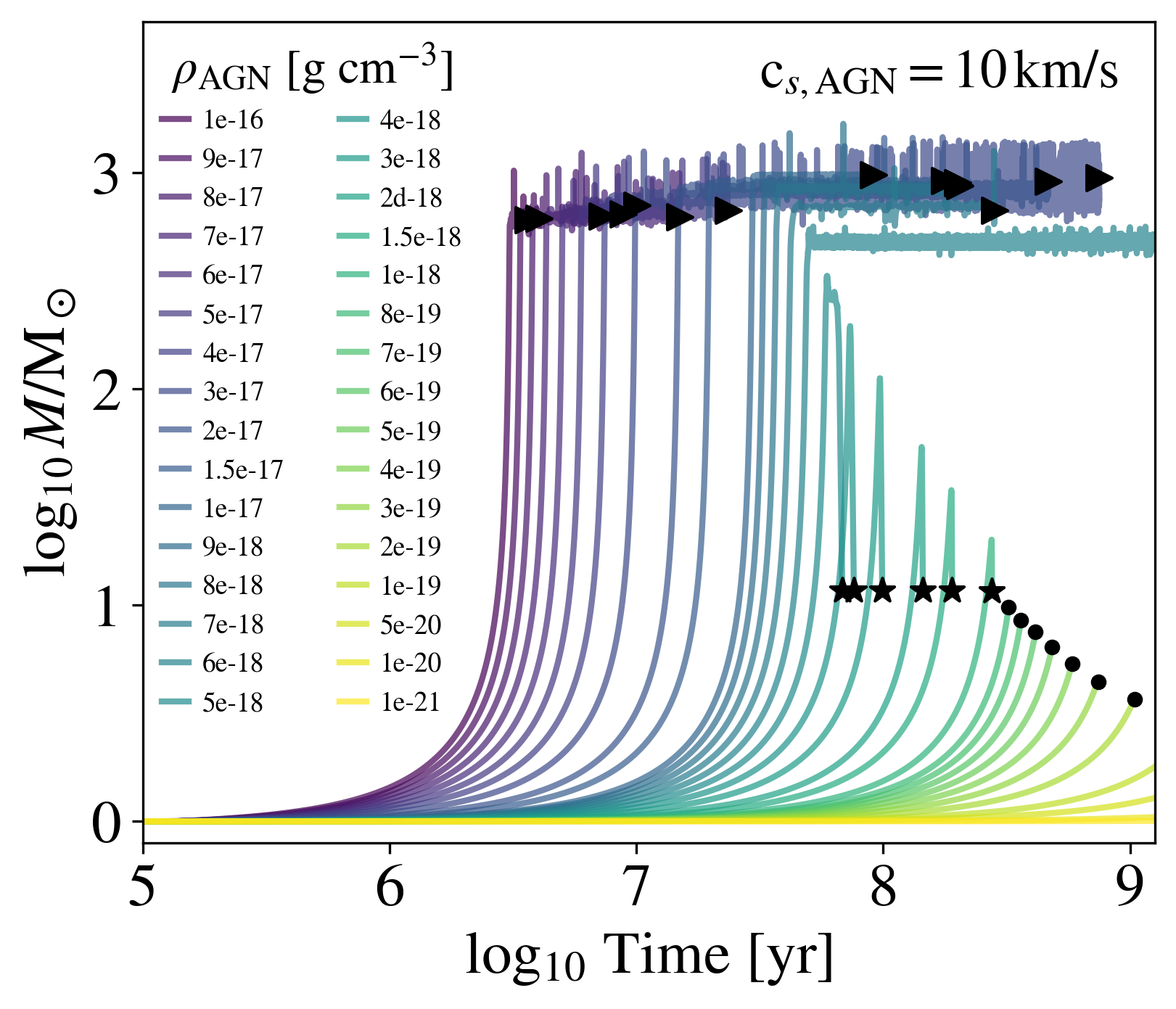}}\hfill
  \subfloat{\includegraphics[width=1.04\columnwidth]{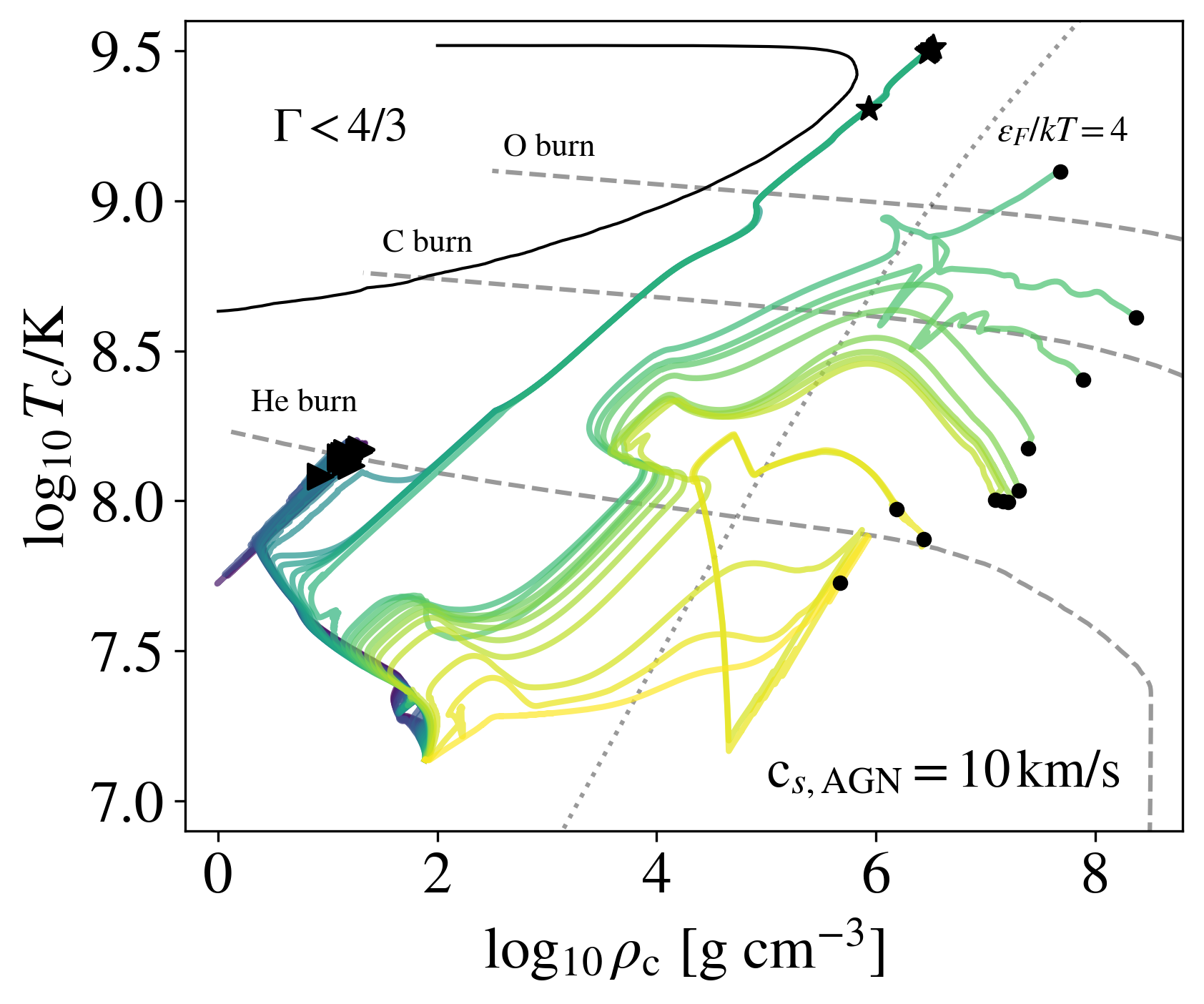}}
   \caption{A grid of stellar models evolved with a fixed AGN sound speed of 10 $\kms$ and AGN densities ranging from $10^{-16}$ to $10^{-21}\,\mathrm{g\,cm^{-3}}$. Left panel: evolution of stellar mass as a function of log time for models starting with $M = 1\mso$. Models evolving at densities higher than $\approx 5\times 10^{-18}\,\mathrm{g\,cm^{-3}}$ experience runaway accretion and become supermassive stars. Models evolving at densities $5\times 10^{-18} \le \rhoagn \le 8\times 10^{-19}$ become massive stars before losing mass via super-Eddington winds and ending their lives with $M \approx 10\mso$ (tracks ending with a star symbol). At densities lower than $\approx 5\times 10^{-19}\,\mathrm{g\,cm^{-3}}$ models end their main sequence evolution before accreting sufficient material to become massive stars ($M \lesssim 8\mso$, tracks ending with a circle). Right panel: Evolution in the central density -- central temperature plane.  Runaway models are continuosly replenished via accretion and burn hydrogen indefinitely (triangles). Intermediate accretion models evolve to late burning stages (tracks ending with star symbols). \agns\ evolution reduces to classic stellar evolution for very small values of $\rhoagn$ (small accretion rates). Dashed lines show the ignition location of different nuclear fuels. The pair instability region is shown as a continuous black line, while the dotted line shows the transition to electron-degenerate gas. }
    \label{fig:gridcs10}%
\end{figure*}

When $\taunuc  \lesssim \tauacc  < \tauagn$ the accretion timescale is comparable to the burning timescale.
\agns\ in these circumstances initially become massive via accretion, but continue to burn hydrogen faster than they accrete fresh fuel.
As a result these stars evolve to late nuclear burning stages.
As they become massive, due to their enhanced internal mixing (see \S~\ref{sec:model}), they also tend to evolve quasi-chemically homogeneously \citep{Maeder:1987}. 
Eventually these \agns\ reach the Eddington luminosity, at which point they tune themselves to sit near $\lstar=\ledd$. 
This happens because on one hand mass loss reduces $\lstar/\mstar$, while on the other hand the chemical evolution of the core serves to increase $\lstar/\mstar$. The gradual increase in the stellar mean molecular weight increases the Eddington ratio $\lstar/\ledd$, so once the star reaches the Eddington limit, it will be forced to constantly lose mass in order to keep near $\lstar = \ledd$ \citep{Owocki:2012}.
During their evolution, these well-mixed stars produce and then expel significant quantities of nuclear ash, which can serve to chemically enrich the AGN disk (\S~\ref{sec:pollution}).
Our intermediate accretion calculations reach late phases of nuclear burning (typically oxygen burning) as H-free, compact stars with $\mstar \approx 10\,\Msun$. These stars are expected to undergo core collapse and produce compact remnants (\S~\ref{sec:remnants}). 
In our model grid with $\csagn = 10\,\kms$, we predict that stars lie in this regime when $ \rhoagn \simeq 5\times 10^{-18}\dots 8 \times 10^{-19}$ g cm$^{-3}$  (tracks ending with a star symbol in Fig.~\ref{fig:gridcs10}).

\begin{figure*}[htp!]
  \centering
  \subfloat{\includegraphics[width=1.05\columnwidth]{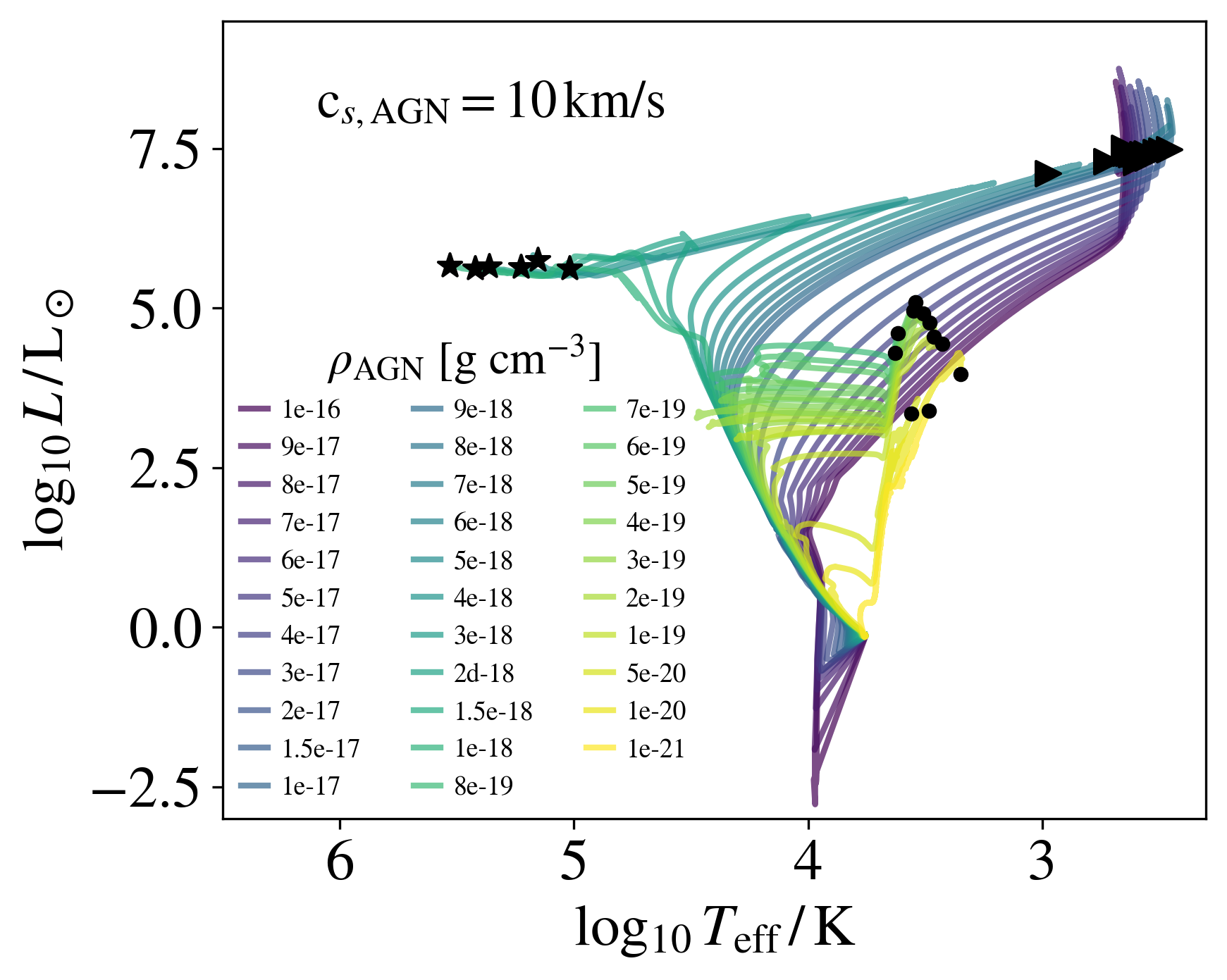}}\hfill
  \subfloat{\includegraphics[width=1.05\columnwidth]{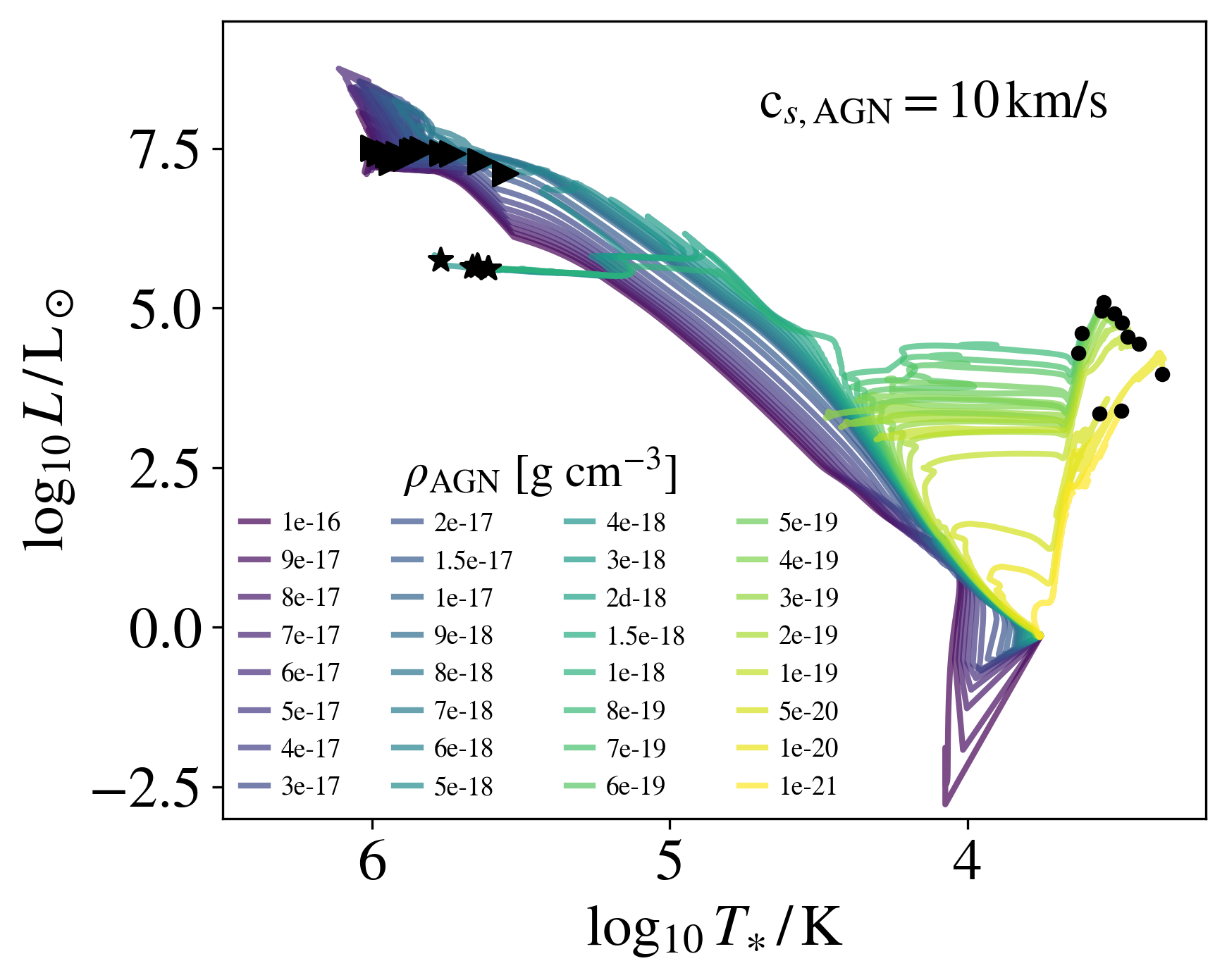}}
   \caption{Evolution on the HR-Diagram for the models shown in Fig.~\ref{fig:gridcs10}. Left panel: The effective temperature of the models is calculated assuming electron scattering opacity for the accretion stream. This is the temperature of the star for an observer sitting at the Bondi radius. Right panel: Same as left panel, but using the surface temperature of the MESA model instead of the effective temperature (Eq.~\ref{eq:Teff}).  This diagram is useful to understand the type of evolution \agns\ models are undergoing, as compared to  canonical stellar evolution.} 
    \label{fig:gridcs10_hrd}%
\end{figure*}

\subsection{Runaway Accretion}
When $\tauacc \ll \taunuc < \tauagn$, \agns\ rapidly become massive.
Due to their enhanced internal mixing (see \S~\ref{sec:model}) these models stay quasi-chemically homogeneous, and hydrogen rich material accreted from the surface is efficiently mixed in the stellar core.
Because $\taunuc \gg \tauacc$ the star is supplied with fresh fuel at a rate faster than it can burn it.
Therefore the evolution stalls on the main sequence, and the runaway accretion process results in a supermassive star ($\mstar \gg 100 \mso$). These stars reach masses of about $1000\,\mso$ before super-Eddington massloss starts balancing accretion. Being so massive, these models are almost fully convective, and stay chemically homogeneous since we assume mixing is very efficient also in radiative regions of stars close to the Eddington limit (Sec.~\ref{sec:mixing}). In this situation the stellar core is continuously supplied of fresh fuel from the AGN disk, and the main sequence lifetime of the model can be extended indefinitely (as long as the accretion rate remains high enough). 

We evolve these models for some time to show this peculiar evolutionary feature of \agns. Some show interesting oscillations of their maximum mass around a mean value which appear stochastic, but since this is likely dependent on the uncertain implementation of mass loss and accretion close to the Eddington limit we did not investigate this feature in depth. Models at the highest range of the accretion rates explored in this work tend to terminate due to numerical reasons when they approach $\approx1000\,\mso$, so it is difficult to predict the final outcome of models at higher densities and/or lower sound speeds. However, it seems plausible that for such conditions \agns\ models would still undergo a runaway accretion phase and reach equilibrium masses $\sim10^3\,\mso$, just on shorter timescales.

We predict \agns\ to lie in this runaway accretion regime when $\rhoagn > 5\times 10^{-18}$ g cm$^{-3}$ and $\csagn = 10\,\kms$ (tracks ending with triangles in Fig.~\ref{fig:gridcs10}), as well as for $\rhoagn > 1.5\times 10^{-19}$ g cm$^{-3}$ in models with $\csagn = 3\,\kms$ (See Fig.~\ref{fig:appendix_gridcs3}). \agns\ models with  $\csagn = 100\,\kms$ require much higher values of density to become supermassive ($\rhoagn > 5 \times 10^{-15}$ g cm$^{-3}$), see  models grid in Fig.~\ref{fig:appendix_gridcs100} and \ref{fig:appendix_gridcs100_hrd}.
For $\csagn = 10\,\kms$ the models that become supermassive  require between $4-50\,\mathrm{Myr}$ to do so. Since $\mdotbondi \propto  \rhoagn \csa^{-3}$, this time decreases substantially for lower values of the AGN sound speed.

If we artificially interrupt accretion, mimicking the star exiting the AGN, entering a gap, or the AGN dissipating altogether (e.g. for $t > \tauagn$), models can evolve to later nuclear burning phases.
This results in the ratio $\lstar/\mstar$ staying close to the Eddington limit, which leads to significant mass loss (Eq.~\ref{eq:medd}). Their evolution is almost identical to what described in \S~\ref{sec:intermediate}, and
the end state for these models is also core-collapse as H-free, compact stars of $\approx10\,\Msun$. In \S~\ref{sec:remnants} we discuss the type of stellar explosions and compact remnants these \agns\ might produce.

Note that \citet{Goodman:2004} and \citet{Dittmann:2020} also reported the possibility of producing massive and supermassive stars in AGN disks. The main difference with our scenario is that they focused on {\it in situ} formation, while we are agnostic to the mechanism producing stellar seeds in AGN disks.  Similar to their study, the maximum mass of our \agns\ is  set by their ability to halt accretion via feedback, although in our case  stars only reach masses of $\approx 1000\,\mso$.

\subsection{Massive and Very Massive Stars in the Inner Regions of AGNs}
Our results show that the inner regions of AGN disks are likely populated by a large number of massive and very massive stars.
The total number of massive stars that an AGN disk can produce via accretion onto low-mass stars formed in-situ or captured from nearby nuclear clusters depends on the AGN properties, in particular its density, sound speed and lifetime. We have shown that for values of the sound speed $\leq 10 \, \kms$, runaway accretion occurs for densities higher than $\approx 10^{-17}$ g cm$^{-3}$. In this case \agns\ reach masses above 100~$\Msun$ in less than 10 Myr, with the accretion timescale decreasing rapidly for higher values of the density and lower values of the local sound speed. 

\begin{figure}[ht!]
\begin{center}
\includegraphics[width=1.0\columnwidth]{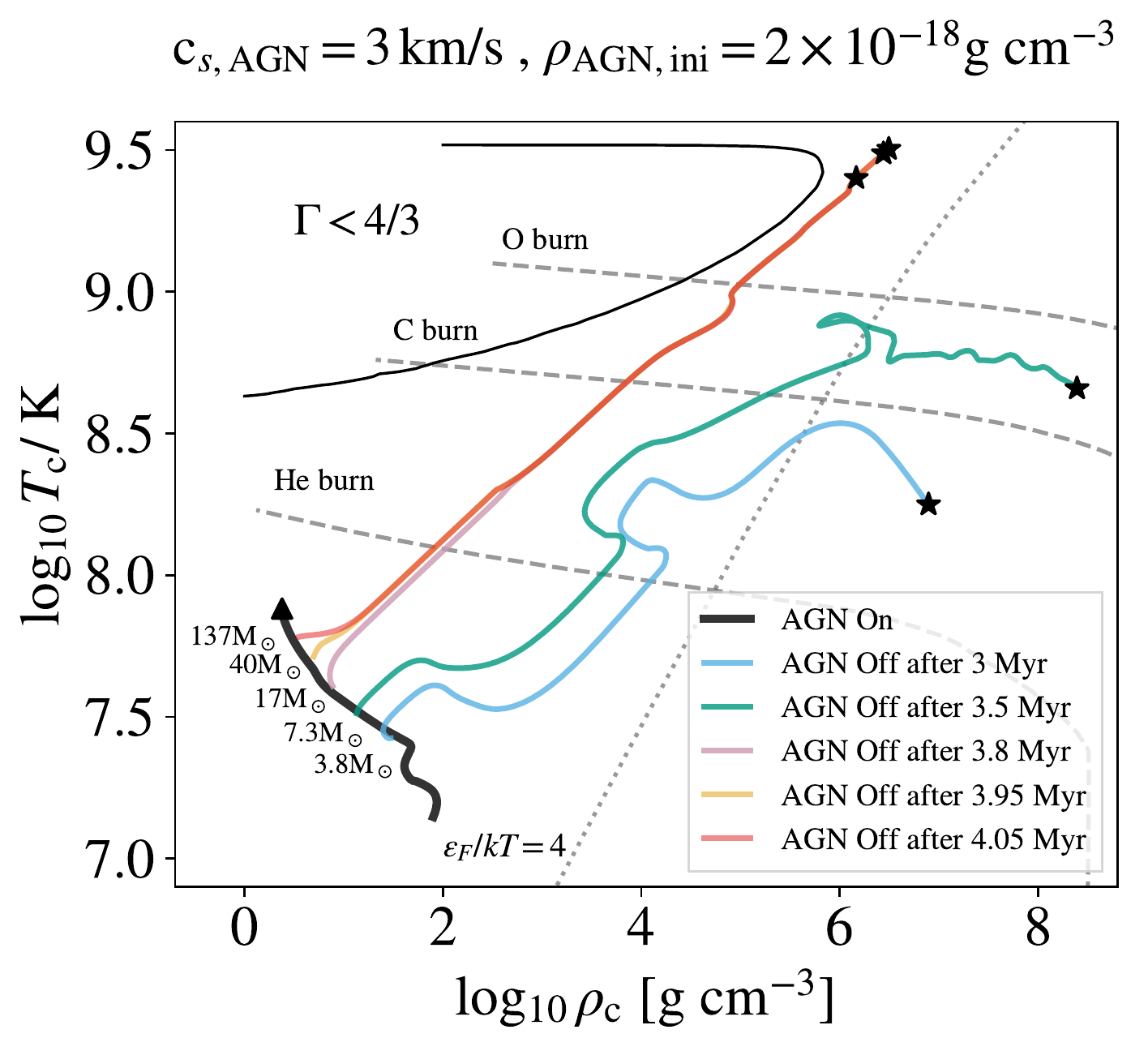}
\caption{\label{fig:agn_shutoff} Evolution of AGN star models in the central density -- central temperature plane for different AGN lifetimes. A $1\,\mso$ model  accretes rapidly at $\rhoagn = 2 \times 10^{-18}\,\mathrm{g\,cm^{-3}}$ and $\csa = 3\,\kms$, becoming a $785\,\mso$ supermassive star in about 4.1 Myr. Halting the accretion before this point, results in different evolutionary pathways. If the initial AGN conditions persist for more than $\approx3.5\,$Myr, the models produce massive stars that undergo core-collapse.}
\end{center}
\end{figure}

\begin{figure}[ht!]
\begin{center}
\includegraphics[width=1.0\columnwidth]{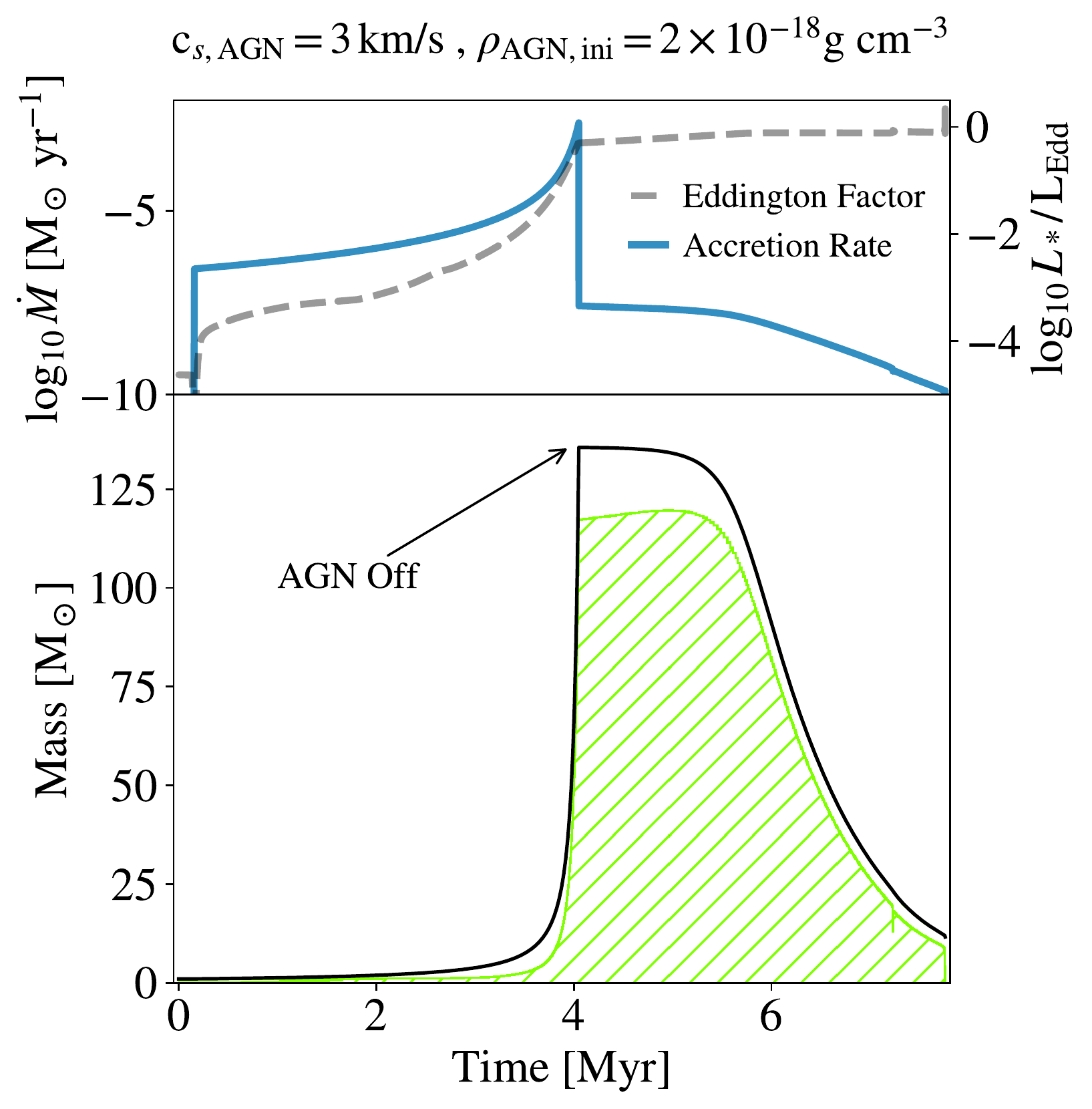}
\caption{\label{fig:kipp_agn_shutoff}  A $1 \mso$ model  accretes rapidly at $\rhoagn = 2\times  10^{-18}\,\mathrm{g\,cm^{-3}}$ and $\csa = 3\,\kms$. We simulate the shutoff of the AGN (or the migration of the AGN star into a gap) by decreasing the local density by a factor $10^5$ after 4.05Myr.  The model stops accreting substantial amounts of mass and evolves past the main sequence (red curve in Fig.~\ref{fig:agn_shutoff}). Similarly to intermediate accretion models, the luminosity stays close to the Eddington limit for most of the subsequent evolution, resulting in large values of the mass loss rate. After about 7 Myr the model ends its life  as a compact He star of $\approx 10\Msun$. Hatched green regions in the lower panel are convective.}
\end{center}
\end{figure}

It is conceivable that accretion is eventually halted via some feedback process, or the star entering regions of the disk with much lower values of the density (or much larger values of the sound speed), or the AGN shutting off altogether. We attempted simulating this occurrence by suddenly decreasing the local density by a factor of $10^5$ for a model of an AGN star initially evolved at $\rhoagn = 2\times10^{-18}\,\mathrm{g\,cm^{-3}}$ and $\csa=3\,\kms$. In this case the evolution of the stellar model  proceeds to later nuclear burning stages, and a large fraction of accreted material is removed via a super-Eddington wind (Fig.~\ref{fig:agn_shutoff} and \ref{fig:kipp_agn_shutoff}). This material has been processed by high temperature nuclear burning, and so it is returned to the AGN highly enriched in helium and metals (Fig.~\ref{fig:AGNShutoff_cumulative_yields_4d-18}). We discuss in \S~\ref{sec:pollution} the implications of the presence of such thermonuclear factories in AGN disks. These massive \agns\ are also expected to leave behind a large population of compact remnants in the inner regions of AGN disks, and we discuss possible observational consequences in \S~\ref{sec:remnants} and \S~\ref{sec:gw}.

\section{Observational Signatures}\label{sec:signatures}
Here we discuss possible observational signatures of \agns\ evolution. We focus on the predicted chemical enrichment of the host AGN, the type of stellar populations that might be left in galactic centers, as well as the stellar explosions and compact remnants resulting from this exotic stellar evolution pathway. Compact remnants in AGN disks are particularly interesting as progenitors of gravitational waves sources, and we briefly discuss the impact of our scenario for this production channel.

\subsection{AGN Pollution}\label{sec:pollution}
While measurements of abundances in AGNs are very difficult and depend on detailed model assumptions \citep{Maiolino:2019},
the consensus is that both high- and low-redshift AGNs appear to
have solar-to-supersolar metallicities \citep[e.g.][]{StorchiBergmann:1989,Hamann:2002,Jiang:2018,Maiolino:2019}.
Despite careful and extensive searches, low-metallicity AGNs seem to be extremely rare \citep{Groves:2006}.

\begin{figure}[ht!]
\begin{center}
\includegraphics[width=1.0\columnwidth]{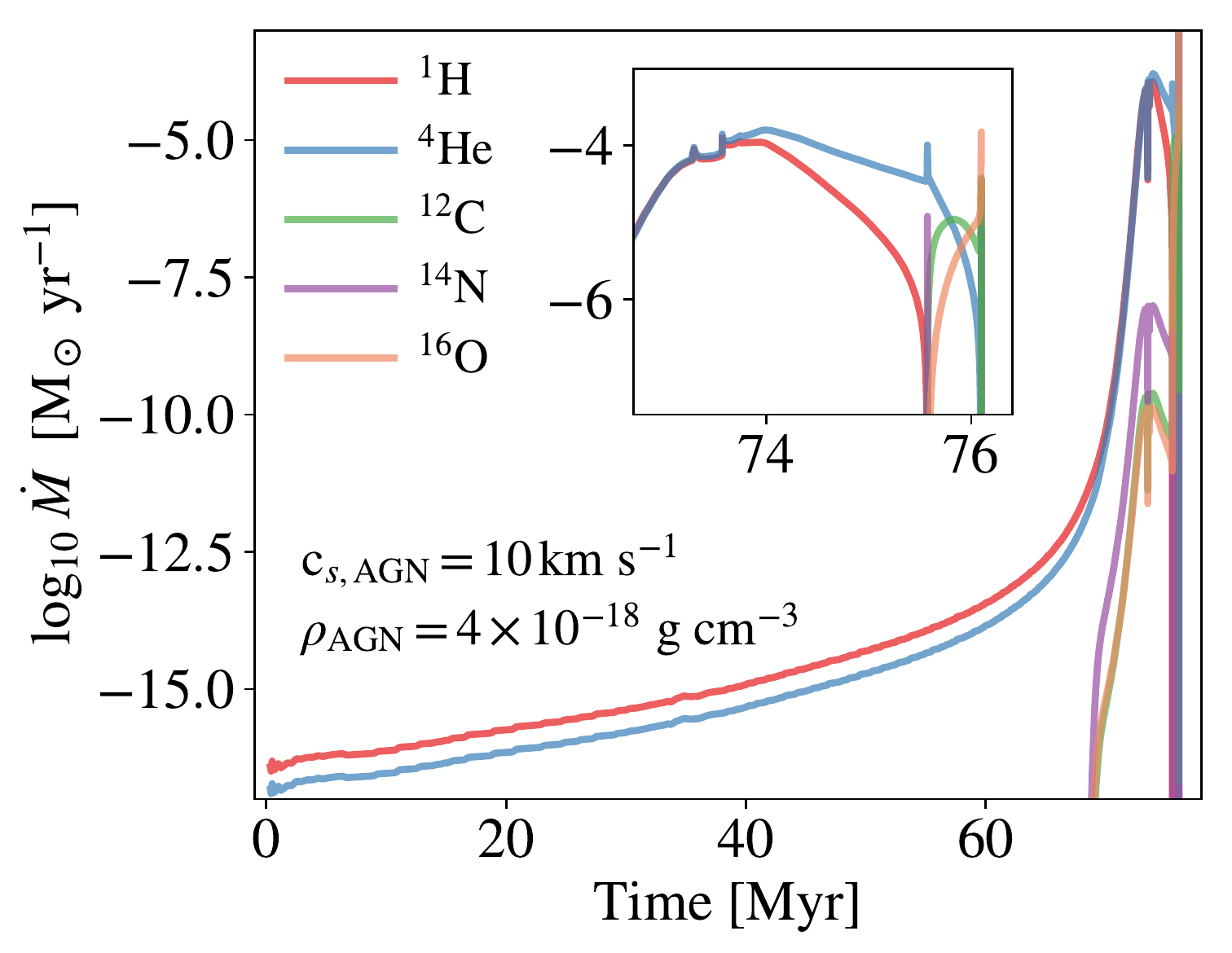}
\caption{\label{fig:yields_4d-18} The mass lost by an AGN star model is shown broken down by species as a function of time. This model was evolved at a density of $4\times10^{-18}$ g cm$^{-3}$ and a temperature of about $186\,\mathrm{K}$ (AGN sound speed of 10 $\kms$). The first part of the evolution is dominated by accretion, and material is lost from the star at a very small rate. As the star approaches the Eddington limit, enhanced mixing dredges processed material up from the core and super-Eddington winds then eject large amount of helium-rich material. As the star evolves to late burning stages, it also loses material rich in carbon and oxygen. 
}
\end{center}
\end{figure}

\begin{figure}[ht!]
\begin{center}
\includegraphics[width=1.0\columnwidth]{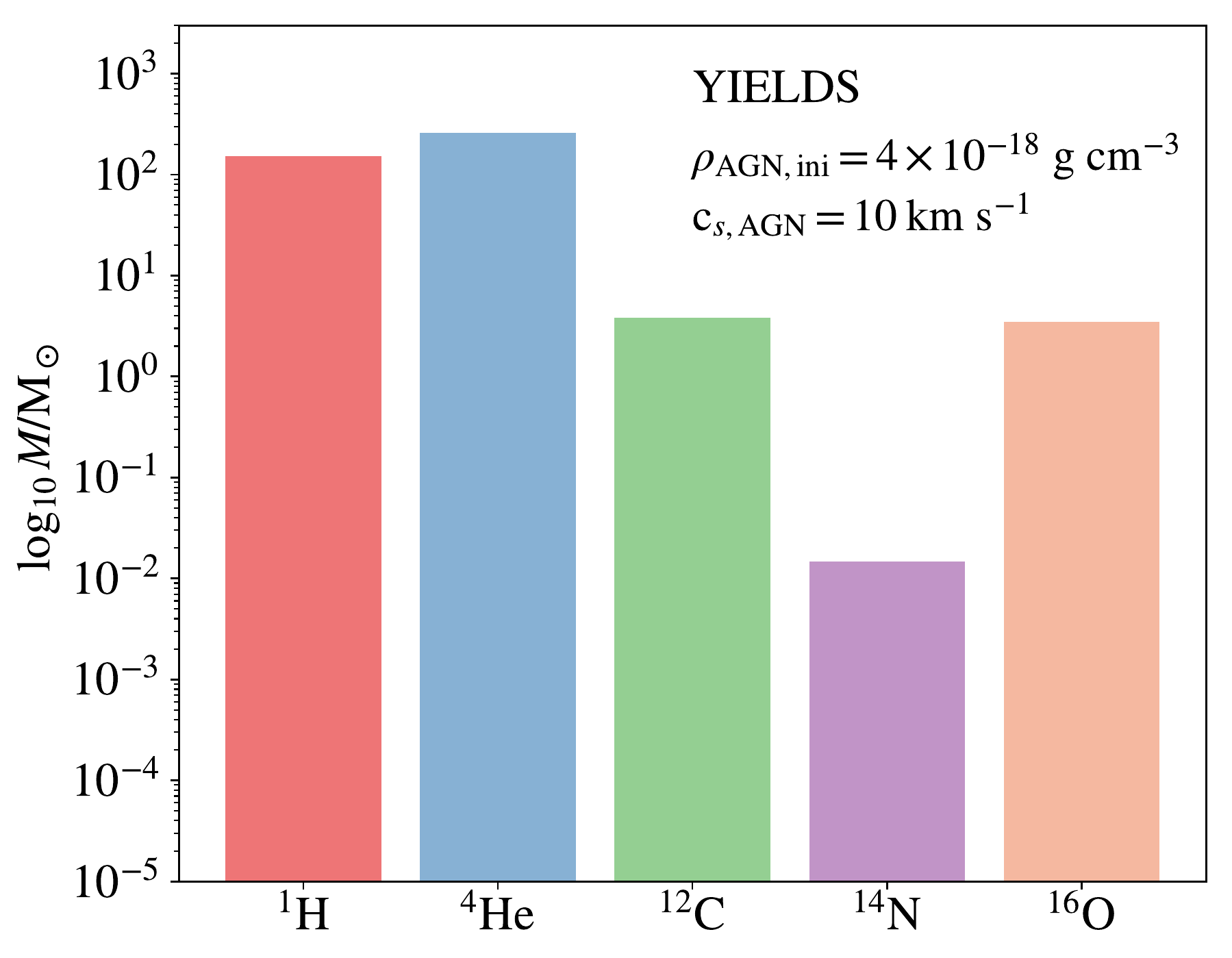}
\caption{\label{fig:cumulative_yields_4d-18} Cumulative yields corresponding to the mass lost by the intermediate accretion model in Fig~\ref{fig:yields_4d-18}. Through its lifetime ($\approx 77$ Myr), this model releases in the AGN 
153.4 $\mso$ of H, 258.6 $\mso$ of He, 3.8 $\mso$ of C, 0.015 $\mso$ of N, and 3.47 $\mso$ of O. This is just accounting for stellar mass loss. A SN explosion would enhance these yields further, and would do so mostly for heavier elements.
}
\end{center}
\end{figure}

\agns\ provide a potential explanation for this observational puzzle. Because  they tend to be longer-lived  (due to rejuvination) and more efficiently mixed than ordinary massive stars, \agns\ are able to process more hydrogen and helium into metals. Moreover, they lose large amounts of mass due to the interplay of accretion and mass loss occurring when they reach their Eddington luminosity (Fig.~\ref{fig:gain_loss_4d-18}). The composition of the material lost by an AGN star model as a function of time is shown in  Fig.~\ref{fig:yields_4d-18}.  This model was evolved till oxygen burning.
While the material which falls onto the star is $72\,\%$ H and $28\,\%$ He, the material which is lost to the AGN disk is 
$36.6\,\%$~H, $61.7\,\%$~He, and $1.7\,\%$ metals by mass (see cumulative yields in Fig.~\ref{fig:cumulative_yields_4d-18}).
Most of the metals are produced towards the end of this model's evolution, and real \agns\ may provide further AGN enrichment via SN explosions and gamma-ray bursts (GRBs).  We tested the effect of adopting different mixing efficiencies on the yields by choosing different values for $\xi$ in Eq.~\ref{eq:mixing} ($\xi$ = [2,4,7]), and found differences of less than 2\%.

\begin{figure}[ht!]
\begin{center}
\includegraphics[width=1.0\columnwidth]{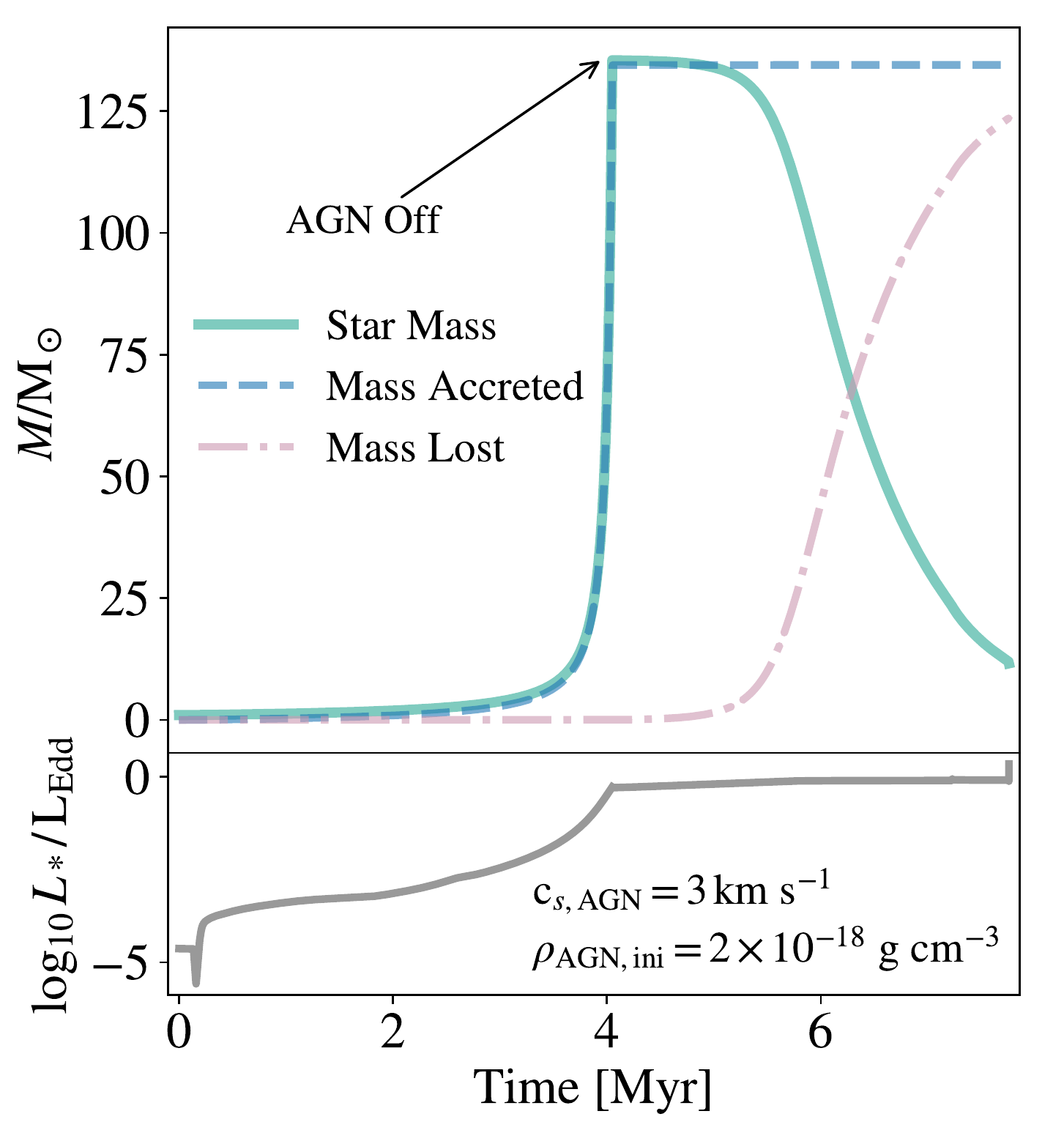}
\caption{\label{fig:AGNShutoff_budget}  Mass budget for the model in Fig~\ref{fig:kipp_agn_shutoff}.  Through its lifetime ($\approx$ 7 Myr), this model accreted from the AGN disk about 134 $\mso$ of gas and released in the AGN approximately 125 $\mso$ of nuclearly-processed material.}
\end{center}
\end{figure}

\begin{figure}[ht!]
\begin{center}
\includegraphics[width=1.0\columnwidth]{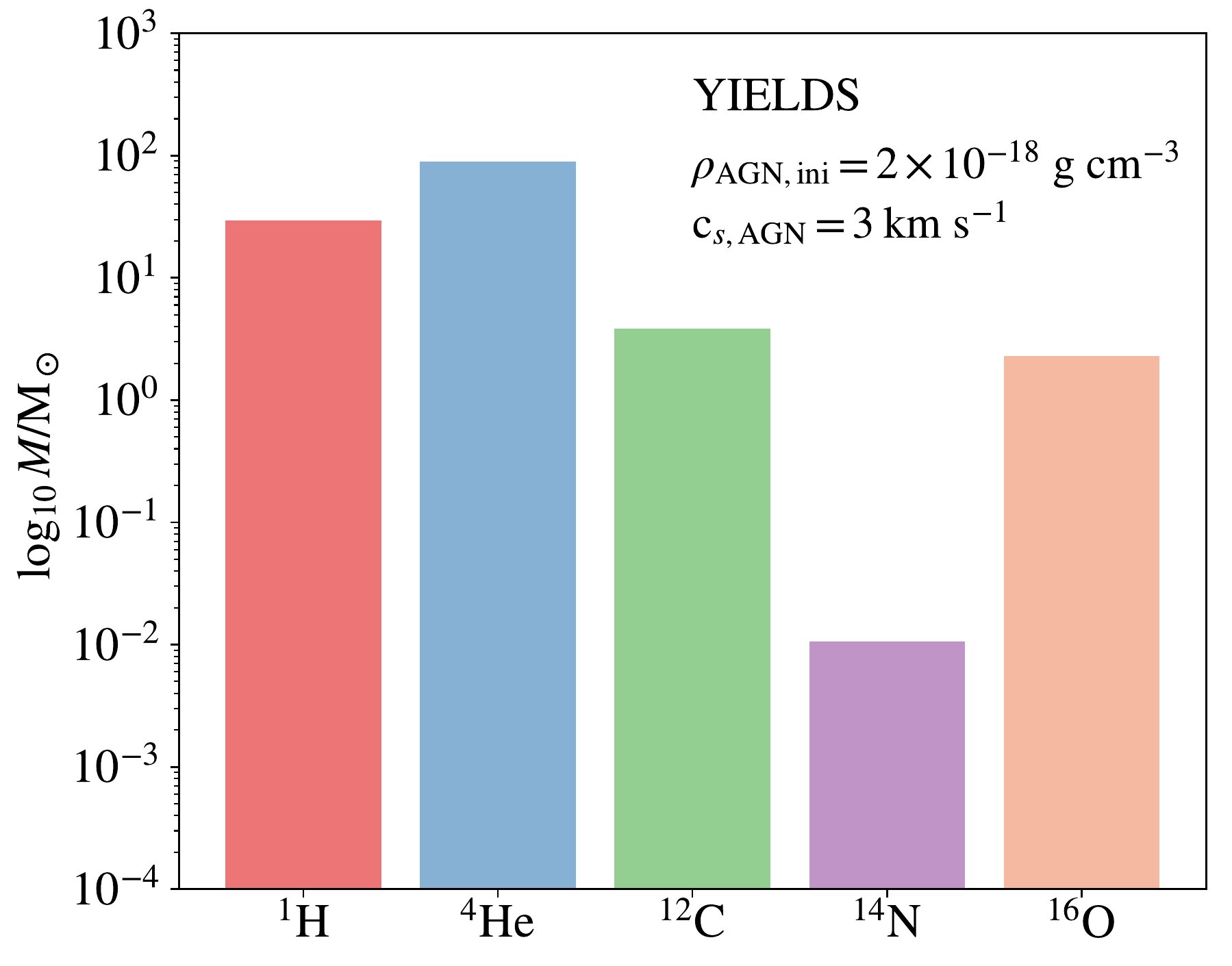}
\caption{\label{fig:AGNShutoff_cumulative_yields_4d-18} Cumulative yields corresponding to the mass lost by the model in Fig~\ref{fig:kipp_agn_shutoff}. This is a runaway accretion model that experienced a shutoff of the AGN (or entered a gap) after 4.05 Myr. Through its lifetime ($\approx$ 7 Myr), this model releases in the AGN 29.6 $\mso$ of H, 89.2 $\mso$ of He, 3.84 $\mso$ of C, 0.011 $\mso$ of N, and 2.29 $\mso$ of O. This is just accounting for stellar mass loss. A SN explosion would enhance these yields further, and would do so mostly for heavier elements.
}
\end{center}
\end{figure}

Note that \agns\ models evolved at higher accretion rates can pollute the disk on much shorter timescales. For example, models initially experiencing runaway accretion but eventually entering regions of the disk with lower densities and/or sound speed rapidly evolve and lose large quantities of chemically-enriched material. This is the case of the model shown in Fig.~\ref{fig:kipp_agn_shutoff}, which  released about $89.2\,\mso$ of He and several solar masses of metals in less than 7 Myr (Fig.~\ref{fig:AGNShutoff_budget} and \ref{fig:AGNShutoff_cumulative_yields_4d-18}). Finally, we want to stress that the yields shown in Fig.~\ref{fig:yields_4d-18} and \ref{fig:AGNShutoff_cumulative_yields_4d-18} are just typical examples, and \agns\ with different disk conditions can produce different outcomes.

To determine whether or not the helium pollution we predict is significant relative to the overall mass budget of the AGN disk, note that the accretion rate in the disk is given by equation~\eqref{eq:mdotshankar} as $2 m_8 {\rm M}_\odot {\rm yr}^{-1}$.
By contrast, our models show that typical massive AGN star models enhances the primordial material of the AGN disk at an average rate of $10^{-4}\dots10^{-5}\,\msol\,\mathrm{yr}^{-1}$.
This suggests a population of $10^4\dots10^5$ \agns\ would suffice to produce significant helium enrichment in the disk.
This calculation is complicated somewhat by the different outcomes of AGN star evolution in different density and temperature regimes, but provides the order of magnitude population that would be required for the chemical enrichment to be significant.

\subsection{Stellar Explosions and Compact Remnants}\label{sec:remnants}
The main feature of AGN stars evolution is the possibility of growing to large masses via accretion, and evolving towards core collapse.  
We expect the relative yields of core collapse supernova, GRBs, and compact remnants  in the inner regions of AGNs to be much larger than what is anticipated for a stellar population with a standard IMF. The light from some of these explosive events might be detectable even if they occur inside AGN disks \citep{Perna:2020,Zhu:2021}.

In our models the interplay of accretion and mass loss (see \S~\ref{sec:massloss}) results in massive \agns\ evolving such that their ratio $\lstar/\mstar$ approaches and stays at the Eddington limit.  For an initial stellar metallicity of Z=0.02,
we find that \agns\ that accrete large amounts of mass subsequently lose most of their mass before reaching core collapse. The strong mass loss rates and enhanced mixing experienced by these stars cause them to reach core-collapse as compact, H-free stars with $
\mstar \approx 10\mso$. In order to predict their likely outcome after core collapse, we compare with the predictions of \citet{Ertl:2020},
who looked at the explodability of helium stars. The presupernova structures of our models allow us to calculate their compactness parameter \citep[e.g.][]{Connor:2011,Sukhbold:2014}, which we then use to predict if the star produces a successfull explosion and what kind of compact remnant is left behind. Since the nuclear network adopted during late phases of burning can affect the presupernova structure and its core compactness \citep{Farmer:2016}, we calculated a few models adopting a larger nuclear network (\texttt{mesa\_128.net}, which uses 128 isotopes). While a systematic study of the compactness parameter of \agns\ will be performed in a subsequent work, in these models we found compactness parameters above 0.2, which when compared with results of  \citet{Ertl:2020} suggests they should produce a BH. These BHs can further accrete and merge, with important implications for gravitational waves sources, as well as LMXBs/HMXBs and GRBs in the inner regions of galaxies. 
A population of energetic explosions embedded in AGNs  could result in important feedback effects on disk accretion and structure, and possibly explain some AGN variability \citep[e.g.][]{Graham:2017}.

\subsection{Gravitational Waves Sources}\label{sec:gw}
The gravitational waves events detected by LIGO-Virgo might originate from stellar mass binary black hole (BBH) mergers in AGN disks \citep[See e.g.][]{McKernan:2012,McKernan:2014,Bartos:2017,Stone:2017,Graham:2020}. The recent detection of a BBH merger with at least one BH formed in the pair-instability mass gap, supports formation channels requiring (multiple) stellar coalescences or hierarchical mergers of lower-mass black holes in AGNs or star clusters \citep[GW190521,][]{Abbott:2020prl,Abbott:2020apj}.  

For the ``AGN channel'', the expected mergers rate is usually parametrized assuming an initial distribution of BHs and stars surrounding the central SMBH, with the BHs formed either in a Toomre-unstable AGN disk or in a nuclear stellar cluster  \citep{Stone:2017,McKernan:2018,Fragione:2019,Tagawa:2020}.  In our scenario, most compact remnants descend from initially low-mass stars that are either captured or formed by the AGN disk, and in regions where the accretion rate is sufficiently high become massive in a timescale shorter than typical AGN lifetimes \citep[1-100 Myr, e.g.][]{Martini:2001,Haiman:2001,Khrykin:2019}.  Regardless of the details of star formation in the regions surrounding SMBH, this process is likely mimicking the effects of having a particularly top-heavy IMF in the inner regions of the AGN disk. Our results show that a large population of compact remnants could be produced within AGNs in regions of the disk where the accretion rate is sufficiently high. 

The evolution of these compact remnants in AGNs is dominated by a number of processes: disk migration and binary formation \citep{Bellovary:2016,Secunda:2019,Tagawa:2020}, hardening via three-body scattering and gaseous drag \citep{Stone:2017,Leigh:2018}, gas accretion and mergers \citep{Yang:2020}, disc-binary interaction \citep{Groebner:2020,Ishibashi:2020}. While the details are complex, the literature supports the possibility of forming larger and larger BH pairs via hierarchical merger and accretion of a seed BH population, eventually allowing the production of intermediate mass black holes (IMBH) in AGNs.
Our work strengthens this suggestion by providing a mechanism to rapidly and efficiently populate the inner regions of AGNs with compact remnants. 

\subsection{The Galactic Center}
An interesting application of \agns\ evolution is the ability to constrain AGN disk parameters and physics using some of the predictions on the type of remnants, stellar explosions and GW sources expected from this peculiar stellar evolution channel. While the space parameter of AGN lifetime, density and sound speed is degenerate, it is likely that progress could be made by pairing different observational proxies. This route could be particularly interesting for studying the Galactic Center (GC), where observations of stellar populations and stellar remnants are directly available.

There is strong evidence for the existence of a central massive black hole of $\approx 4\times 10^6 \mso$ in the Milky Way \citep{Genzel:2010}. While the accretion rate onto this SMBH is currently low, observations suggest that \sgra\ was much brighter in the recent past \citep{Su2010}. If the Milky Way experienced an AGN phase not too long ago, the properties of the stellar populations and remnants currently observed in GC can be used to test some of the ideas discussed in this paper. 

The central parsec contains about 200 young massive stars. The presence of so many young stars in the immediate vicinity of the central black hole is unexpected \citep{Ghez:2003,Alexander:2005}. The surface density of a group of massive O/WR stars in the inner 1pc region raises steeply from $\sim$13 arcseconds (0.5pc) to a few arcseconds, with no O/WR stars observed beyond 0.5pc \citep{Paumard:2006,Bartko:2010}. Another group of early-type B stars (S-star cluster) shows a similar sharp inward increase of its surface density \citep{Bartko:2010}. Stellar spectroscopy shows that some of these stars might be He-rich \citep{Martins:2008,Habibi:2017,Do:2018}.
There is evidence that the present day stellar mass function (PMF) for the group of centrally-concentrated O/WR stars is flat, and it becomes steeper moving further out \citep{Paumard:2006,Bartko:2010}. The agreement is that the PMF within 0.5pc is top-heavy \citep{Genzel:2010}.   

The two-body relaxation-time scale in the central parsec ranges between 1 and 20 Gyr, much longer than the lifetime of the B and O/WR stars \citep{Alexander:2005}, so this central concentration of massive stars can not be a Bahcall-Wolf cusp \citep{Bahcall:1976,Bahcall:1977}. On the other hand,  lower mass old stars do not exhibit this central concentration, their distribution flattening close to the GC \citep{Genzel:2010,Do:2017}. This is  contrary to basic theoretical predictions \citep{Bahcall:1976,Bahcall:1977}. A number of mechanisms have been proposed to account for the anomalous properties of the stellar population in the GC, including in-situ star formation and in-spiral of a star cluster \citep[See][for a review of these scenarios and their challenges]{Genzel:2010}. In the context of \agns\ evolution, the observations could be explained by low-mass stars accreting large amounts of mass in the inner $\approx 1$ pc region of an AGN disk some $\sim$6 Myr ago. This could naturally account for the surface densities of the different stellar populations, a radially-dependent top-heavy PMF, and chemical peculiarities observed in some of the B-stars spectra.

The low-mass X-ray binary (LMXB) candidates identified in the galactic center by \citet{Hailey:2018} are found only within $\approx 1$ pc. A possible explanation of this peculiar distribution relies on the migration of compact remnants binaries formed via gas-capture mechanism during a former AGN phase \citep{Tagawa:2020}. However, in the context of \agns\ evolution, the 1~pc cutoff could simply represent the radial distance beyond which the gas conditions did not allow for enough accretion to form massive and very massive stars. Beyond this radius, stars can still accrete substantial amounts of mass, but they did not evolve towards core collapse. Owing to their longer evolutionary timescales, some of these intermediate mass objects could still be present in the GC, possibly showing spectroscopic signatures of accreted AGN material \citep{Martins:2007,Cunha:2007,Martins:2008,Habibi:2017,Do:2018} (see \S~\ref{sec:pollution}). 

\section{Conclusions}
Stellar evolution in AGN disks proceeds quite differently than in a vacuum. 
Depending on the density and sound speed of the gas in which they are embedded, abundant low-mass stellar seeds provided either via capture or in-situ formation can rapidly accrete and become massive or super-massive stars. These stars have large convective cores and their luminosity approaches the Eddington limit, so their interiors are prone to mixing. The interplay between accretion and mass loss is likely to enforce envelope circulations, with the mixing timescale decreasing rapidly as the star grows in mass. Therefore, we expect \agns\ that accrete substantial amounts of mass to evolve quasi-chemically homogeneously \citep{Maeder:1987,Yoon:2005}, their surface composition largely reflecting the composition of their nuclear-burning cores.  \agns\ evolving as massive and very-massive stars are expected to lose large amounts of mass enriched in helium and CNO-processed elements. When evolving to core-collapse, they can also expel large amounts of heavy elements via SN explosions, further polluting their AGN disk. Overall the metallicity of AGNs is expected to increase rapidly in the presence of \agns.

Due to the efficiency of internal mixing, \agns\ that become massive and very massive  are only able to evolve when the accretion rate timescale becomes comparable or longer than the nuclear burning timescale. Depending on the value of density and sound speed of the medium they are embedded, our \agns\ models are in this regime either from the beginning of the calculation, or they enter it when the accretion rate decrease substantially due to a change in the AGN disk conditions. This could be due to the end of the AGN phase, or the star migrating to regions of the disk with different gas conditions (e.g. gaps). Our AGN star models that accreted large amounts of mass eventually end their evolution as compact, H-free stars of about $10~\mso$ that undergo core collapse. This outcome is caused by the interplay of accretion and mass loss, which in our implementation keeps these chemically homogeneous stars evolving near $\lstar=\ledd$. Therefore, we expect \agns\ to efficiently populate AGN disks with compact remnants, to a rate much higher than what expected from a population of stars with a canonical IMF. While we defer a thorough study of the parameter space of \agns\ to future works, the models computed for this paper point to an efficient production of compact remnants in the inner regions of AGN disks. These compact remnants are interesting seeds for the growth of BH via further accretion and/or mergers, with important implications for the gravitational waves sources observed by LIGO-Virgo, and for LISA predictions. A population of very massive stars and energetic explosions in the inner regions of galaxies could also have important consequences for the structure and accretion properties of AGNs.

\acknowledgments
We thank Alexander Dittmann, Saavik Ford, Yan-Fei Jiang, Yuri Levin, Barry McKernan, and Mathieu Renzo for useful discussions. We thank the anonymous reviewer, whose comments and suggestions helped improve and clarify this manuscript. The Center for Computational Astrophysics at the Flatiron Institute is supported by the Simons Foundation. This research was supported in part by the National Science Foundation under Grant No. NSF PHY-1748958 and by the Gordon and Betty Moore Foundation through Grant GBMF7392.

\software{%
  \texttt{MESA} \citep{Paxton:2011, Paxton:2013, Paxton:2015, Paxton:2018, Paxton:2019},
  \texttt{ipython/jupyter} \citep{PER-GRA:2007,Kluyver:2016aa},
  \texttt{matplotlib} \citep{Hunter:2007},
  \texttt{NumPy} \citep{harris2020array}, and
  \texttt{Python} from \href{https://www.python.org}{python.org}.
}

\appendix

\section{Temperature and density range in AGN disks}
\label{sec:alphadisk}
We adopt the conventional $\alpha$-disk model \citep{Shakura1973} for effective viscosity 
$\nu = \alpha h^2 \Omega R^2$, where $h=H/R$ is the aspect ratio, $\alpha$ is an efficiency
factor for turbulent viscosity, and $\Omega (=G M_{bh}/R^3)^{1/2}$ is the angular frequency at a 
distance $R$ from the SMBH. In a hydrostatic equilibrium,  the disk scale height is
\begin{align}
    H = \frac{\sqrt{2}}{\Omega}\csa\simeq \frac{\sqrt{2}}{\Omega}\sqrt{\frac{R_g T_{\rm AGN, gas}}{\mu}}
\label{eq:Hgas}
\end{align}
in the gas-pressure dominated region and
\begin{align}
    H = \frac{\sqrt{2}}{\Omega}\csa\simeq  \frac{4 a T_{\rm AGN, rad}^4}{3\Sigma \Omega^2}
\label{eq:Hrad}
\end{align}
in the radiation-pressure dominated region. 
In the above expressions,  $T_{\rm AGN, gas}$ (or $T_{\rm AGN, rad}$) is the mid-plane temperature, $\csa$ is 
the mid-plane sound speed, and $\Sigma$ is the surface density. 
The relative importance of radiation pressure $P_{\rm rad}$ compared to gas pressure $P_{\rm gas}$ is measured by
\begin{equation}
    \beta_P = \frac{P_{\rm gas}}{P_{\rm rad}} = \frac{3 R_g \rhoagn}{\mu a \tagn^3},  
    \label{eq:betap}
\end{equation}
which is generally smaller than unity in the outer regions of the disk.   

In a steady state,  the mass flux through the disk
\begin{equation}
{\dot M}_d = 3 \pi \Sigma \nu \simeq 6 \pi \alpha \rhoagn h^3 \Omega R^3 
\label{eq:mdot}
\end{equation}
where $\Sigma=2 \rhoagn H$ and $\rhoagn$ are the surface density and mid-plane mass density.
We use the most probable values of $\lambda \,(\sim 0.6)$ and $\epsilon \,(\sim 0.06)$ for the Eddington factor and 
the mass-to-energy conversion efficiency 
respectively (see \S~\ref{sec:introduction}). These are obtained from AGN evolution models \citep{Shankar2009}. We use these values to 
derive a ${\dot M}_d-M_{bh}$ relationship and remove one degree of freedom for the input model parameters.  
We scale the SMBH's mass by $m_8 \equiv M_{bh}/10^8 \Msun$ and distance in the disk from it by $r_{pc} \equiv R/{\rm 1 pc}$ 
so that the Keplerian speed is $V_k \simeq 7 \times 10^2 m_8^{1/2} r_{pc}^{-1/2}\, $km s$^{-1}$, the angular frequency
$\Omega = 2.1 \times 10^{-11} m_8^{1/2} r_{pc}^{-3/2}$ \, s$^{-1}$, and   the rate of mass accretion onto the SMBH
\begin{equation}
    {\dot M}_{bh} \simeq {\lambda f_m \over \epsilon} {\ledd \over C^2} \simeq 2 f_m m_8 \,{\rm M}_\odot \,{\rm yr}^{-1},
    \label{eq:mdotshankar}
\end{equation}
where the factor $f_m \,(\sim 1)$ takes into account the dispersion in both $\lambda$ and $\epsilon$. 

In conventional steady-state accretion disk models \citep{Shakura1973, LyndenBell1974}, 
  ${\dot M}_d \simeq  {\dot M}_{bh}$  and the 
viscous dissipation provides a heating rate  
\begin{equation}
    Q^+= 9 \Sigma \nu \Omega^2/4 = 3 {\dot M}_d \Omega^2 / 4 \pi.
\end{equation}
The cooling rate due to radiative diffusion in the direction normal to the disk is
\begin{equation}
Q^- = 2 \sigma T_e^4 = 2 \sigma \tagn^4 \tau/(1+\tau^2),
\label{eq:qminus}
\end{equation}
where $T_{\rm c}$ and $T_e$ are the midplane and surface effective 
temperatures respectively, and $\tau=\kappa \Sigma/2$ is the optical depth.  
In low-density AGN disks, the dominant source of opacity $\kappa$ above $\sim 2 \times 10^3$ K is  electron scattering $\kappa_{\rm es} = 0.2 (1 + x)$ cm$^2$ g$^{-1}$, where $x$ is the ionization fraction \citep[but see also][]{Jiang:2016}.  
Below the sublimation temperature of the refractory grains, dust opacity dominates  and $\kappa_{\rm dust} \sim 0.1\, \tagn ^{1/2} 
10^{[{\rm Fe/H}]}\mathrm{cm^2\,g^{-1}}$, where $[{\rm Fe/H}]$ is the metallicity of the gas relative to the solar value \citep{Bell1994}.  
For convenience, we approximate both $\kappa_{\rm es} \, (\simeq 0.4$ cm$^2$ g$^{-1}$) and 
$\kappa_{\rm dust} \, (\sim 4 \times 10^{[{\rm Fe/H}]}$ cm$^2$ g$^{-1}$) as constants.  

In a thermal equilibrium  $Q^+ = Q^-$. Using Eq.~\ref{eq:mdotshankar}, the effective temperature becomes 
\begin{equation}
    T_e = \left( {3 {\dot M}_{bh} \Omega^2 \over 8 \pi \sigma} \right)^{1/4} \simeq 
    110 {f_m ^{1/4} m_8^{1/2} \over r_{pc}^{-3/4}}{\rm K}.
\end{equation}
Equations (\ref{eq:Hgas}) and (\ref{eq:mdot}) lead to midplane temperature, density, and pressure
\begin{align}
    \label{eq:Tagas}
   T_{\rm AGN, gas} &= \left({ \kappa \mu {\dot M}^2 \Omega^3 \over 16 \pi^2 \sigma \alpha R_g} \right)^{1/5}
     =  T_{\rm g} \left( {\kappa \mu \over \alpha} \right) ^{1/5} 
    {  f_m^{2/5}m_8 ^{7/10}  \over r_{\rm pc}^{9/10}}, \\
    \label{eq:rhoagas}
        \rho_{\rm AGN, gas} &= 
    {\mu^{6/5} \Omega^{11/10} \over 3 R_g^{6/5} \alpha^{7/10}}
    \left( {{\dot M}^4 \sigma^3 \over 8 \pi^4 \kappa^3} \right)^{1/10}    
     = { \rho_{\rm g} \mu^{6/5} f_m^{2/5} m_8 ^{19/20} \over \kappa^{3/10} \alpha^{7/10} r_{\rm pc} ^{33/20}},\\
        \label{eq:pgas}
P_{\rm gas} &= \left( {\sigma \over 2 \kappa} \right) ^{1/10} \left( {\mu {\dot M}^2 \over R_g \pi^2 } \right) ^{2/5}
{\Omega^{17/10} \over 6 \alpha ^{9/10}} 
\end{align}
with $T_{\rm g}=3.7 \times 10^3$K and $\rho_{\rm g} = 1.5 \times 10^{-13} {\rm g \ cm^{-3} }$ for the gas 
pressure dominated region and 
\begin{align}
        \label{eq:Tarad}
T_{\rm AGN, rad} &= \left({ c \Omega \over  2 \alpha a \kappa} \right)^{1/4} 
= T_{\rm r}  \left( {m_8 \over \alpha^2 \kappa^2 r_{\rm pc}^3} \right)^{1/8}\\
        \label{eq:rhoarad}
                \rho_{\rm AGN, rad} &= {\pi^2 c^3 \over 3 \alpha \kappa^3 {\dot M}_d^2 \Omega} 
                = {\rho_{\rm r}  r_{\rm pc}^{3/2} \over \alpha \kappa^3 f_m^2 m_8^{5/2}}, \\
        \label{eq:parad}
        P_{\rm rad}  &=  {c \Omega \over 6 \kappa \alpha},
\end{align}
with $T_{\rm r} = 2.6 \times 10^3$K, $\rho_{\rm r}= 2.2 \times 10^{-10} {\rm g \ cm^{-3}} $
for the radiation pressure dominated region.  The solutions provide the $\tagn$ and $\rhoagn$
distribution for different values of $M_{bh}$ and ${\dot M}_d$.  From 
equations~\eqref{eq:betap},\eqref{eq:Tagas}-\eqref{eq:pgas} and \eqref{eq:Tarad}-\eqref{eq:parad}, 
the boundary separating the gas and radiation pressure dominated regions (where $\beta_P =1$) occurs at 
\begin{equation}
\begin{split}
     R_\beta = {(G M_{bh})^{1/3} ( \alpha a) ^{2/21} \over 2^{2/7} (c/\kappa)^{6/7}}
    \left( { \mu {\dot M}_d^2 \over \pi^2 R_g} \right)^{8/21} \\
    =0.09 \mu^{8/21} \kappa^{6/7} \alpha^{2/21} f_m^{16/21} m_8^{23/21} {\rm pc}.
    \end{split}
\end{equation} 
The gas pressure is dominant for $R< R_\beta$ whereas radiation pressure is dominant for $R > R_\beta$.

This steady state geometrically-thin viscous disk model provides an useful estimate for the values of $T_{\rm c}$ and $\rhoagn$ (AGN midplane temperature and density).
However, it is thermally unstable in the radiation pressure dominated region.  Such an instability is incompatible with the 
thermal equilibrium assumption based on which the model is constructed \citep{Pringle1981}. Moreover these solutions 
suggest the possibility of gravitational instability in the outer regions of the disk.

\section{Software Details}
\label{sec:software}

Calculations were done with \code{MESA} version 15140. The \code{inlist} and \code{MESA} extension required to reproduce the results in this work are available at~\url{https://doi.org/10.5281/zenodo.4437705}.
The \code{MESA} EOS is a blend of the OPAL \citet{Rogers2002}, SCVH
\citet{Saumon1995}, FreeEOS \citet{Irwin:2004}, HELM
\citet{Timmes2000}, and PC \citet{Potekhin2010} EOSes.
Radiative opacities are primarily from OPAL \citep{Iglesias1993,
Iglesias1996}, with low-temperature data from \citet{Ferguson2005}
and the high-temperature, Compton-scattering dominated regime by
\citet{Buchler1976}.  Electron conduction opacities are from
\citet{Cassisi2007}.
Nuclear reaction rates are a combination of rates from
NACRE \citep{Angulo1999}, JINA REACLIB \citep{Cyburt2010}, plus
additional tabulated weak reaction rates \citet{Fuller1985, Oda1994,
Langanke2000}. Screening is included via the prescription of \citet{Chugunov2007}.
Thermal neutrino loss rates are from \citet{Itoh1996}.
We adopted a 21-isotopes nuclear network (\code{approx21.net}), except for a few models where we tested the impact of adopting a 128-isotopes network (\code{mesa\_128.net}). 
Convective boundaries were modeled using the Schwarzschild criterion and no overshooting. 

\section{Thermal and Advection Times}
\label{sec:thermal_adv}

Our aim is to show that the thermal time for the accretion stream $t_{\mathrm{th}}$ is short compared with the advection time $t_{\mathrm{adv}}$.
Given that assumption, the full stellar luminosity must escape through the accretion stream, meaning that the stream carries $L=\lstar$.
The ratio between the two time-scales is then
\begin{align}
	\frac{t_{\mathrm{th}}}{t_{\mathrm{adv}}} = \frac{m_{\mathrm{above}} c_\mathrm{p} T / \lstar}{m_{\mathrm{above}}/\dot{M}} = \frac{c_\mathrm{p} T \dot{M}}{\lstar} = \frac{c_\mathrm{p} T \dot{M}}{\ledd}\left(\frac{\ledd}{\lstar}\right),
\end{align}
where $m_{\mathrm{above}}$ is the mass above a given spherical shell and $T$ is the temperature at the surface of the star.
Letting $\kappa$ be the opacity of the accreting material, we write
\begin{align}
	\ledd = \frac{4\pi G \mstar c}{\kappa}
\end{align}
and find with equation~\eqref{eq:mbondi}
\begin{align}
	\frac{t_{\mathrm{th}}}{t_{\mathrm{adv}}} =\frac{c_\mathrm{p} T \kappa \eta \rbondi^2 \rhoagn \csa}{G \mstar c}\left(\frac{\ledd}{\lstar}\right).
\end{align}
Inserting equation~\eqref{eq:rbondi} we find
\begin{align}
	\frac{t_{\mathrm{th}}}{t_{\mathrm{adv}}} =\frac{4 c_\mathrm{p} T \kappa \eta G \mstar \rhoagn}{\csa^3 c}\left(\frac{\ledd}{\lstar}\right).
\end{align}
With $c_{\mathrm{p}} T \approx \cs^2$ and $\eta = 1$, we then write
\begin{align}
	\frac{t_{\mathrm{th}}}{t_{\mathrm{adv}}} =2 \rbondi\ \kappa \rhoagn \left(\frac{\csa}{c}\right)\left(\frac{\cs^2}{\csa^2}\right)\left(\frac{\ledd}{\lstar}\right).
\end{align}
Using $\rbondi \la 10^{15}\cm$, $\rhoagn \la 10^{-14}\mathrm{g\,cm^{-3}}$, $\kappa \la 1\,\mathrm{cm^2\,g^{-1}}$, $\csa/c \la 10^{-3}$, and $\cs / \csa < 10$, we find that this ratio is no more than
\begin{align}
	\frac{t_{\mathrm{th}}}{t_{\mathrm{adv}}} \la \left(\frac{\ledd}{\lstar}\right).
\end{align}
In these extreme cases with large Bondi radii and high AGN densities the luminosity quickly rises to Eddington in our models, so this ratio is at most unity and it is a good approximation to let the entropy of the accreting material equal that of the surface of the model.
In less extreme cases $L$ may be much less than $\ledd$, but then the pre-factor is much smaller and the approximation is again good~\citep{Paxton:2015}.

We could alternatively have assumed that the accretion stream is advection-dominated and derived a contradiction.
This proceeds as follows.
The material which falls in adjusts to its new density adiabatically, following $\rho \propto r^{-3/2}$ (Eq.~\ref{eq:rho_scale}).
The radiative luminosity the material carries is then given by equation~\eqref{eq:dTdr} as
\begin{equation}
    L = -\frac{64 \pi r^2 \sigma T^3}{3 \kappa \rho}\frac{dT}{dr},
\end{equation}
which implies a thermal time-to-advection time ratio of
\begin{equation}
    \frac{t_{\rm th}}{t_{\rm adv}} = \frac{3 \kappa \rho c_p T \dot{M}}{64 \pi r^2 \sigma T^3}\frac{dr}{dT}.
\end{equation}
For an ideal gas with $\gamma=5/3$, the infalling material has a temperature gradient
\begin{equation}
    \frac{dT}{dr} \propto \frac{d(P/\rho)}{dr} \propto \frac{d(\rho^{\gamma-1})}{dr} \propto r^{-(3/2)(\gamma-1)-1} = r^{-7/2},
\end{equation}
so
\begin{equation}
    \frac{t_{\rm th}}{t_{\rm adv}} = \frac{3 \kappa \rho c_p T \dot{M}}{224 \pi r \sigma T^4}.
\end{equation}
This expression scales as $\rho r^{-1} T^{-3} \propto r^8$, so it is maximized at large $r$.
Evaluating this at the Bondi radius and expanding $\dot{M}$ with equation~\eqref{eq:mbondi} we find
\begin{equation}
    \frac{t_{\rm th}}{t_{\rm adv}} = \frac{3 \kappa \rhoagn^2 \csagn c_p \tagn \eta \rbondi }{224 \sigma \tagn^4}.
\end{equation}
Using $c_p T \approx \csagn^2$ and equation~\eqref{eq:rbondi} we then obtain
\begin{equation}
    \frac{t_{\rm th}}{t_{\rm adv}} = \frac{3 \kappa \rhoagn^2 \csagn \eta G \mstar }{112 \sigma \tagn^4} = 6\times 10^{-11} \eta \left(\frac{\kappa}{10^{-3}\mathrm{cm^2\,g^{-1}}}\right)\left(\frac{\mstar}{\Msun}\right)\left(\frac{\tagn}{10^3\mathrm{K}}\right)^{-4}\left(\frac{\csagn}{10^6\mathrm{cm\,s^{-1}}}\right)\left(\frac{\rhoagn}{10^{-15}\mathrm{g\,cm^{-3}}}\right)^{2}.
\end{equation}
Because this is extremely small, the thermal time is much shorter than the advection time for most choices of parameters.

\section{Accretion Stream Assumptions}\label{sec:assumptions}

We now examine the assumptions we made in determining the properties of the accretion stream.
Our aim is not to ensure that every assumption is exactly upheld, but rather to see whether, if they fail to hold, that makes an order-of-magnitude or scaling difference to our results.

We begin with spherical symmetry and the steady state assumption.
Both of these assumptions are almost certainly wrong: the accretion stream is neither spherically symmetric nor time-independent.
Rather there will be regions that are over-dense and under-dense, or moving faster or slower, and which regions these are may vary with time.

We have attempted to incorporate these effects qualitatively in our treatments of mass loss and gain, because there the structure of the inflows and outflows likely matters and there are clear geometric effects such as that an inflow and outflow cannot occupy the same space.

For the boundary conditions on temperature and pressure we believe these assumptions matter much less.
For temperature fluctuations this is because changes on a scale less than a thermal time of the accretion stream do not alter the structure of the star or its nuclear burning.
This is because the fluctuation time for the stream is of order the free-fall time from $\rbondi$ to $\rstar$, which is short compared with the thermal time-scale for most of the star.
For pressure fluctuations we already incorporated the effects of fluctuations in \S~\ref{sec:mixing} as an overall enhancement in mixing, as the main effect of these is to produce waves and flows which may mix the star.
The same is true of aspherical temperature and pressure perturbations: these mostly serve to induce mixing~\citep[e.g.][]{1925Obs....48...73E}  and typically become less important the further one looks in the star~\citep[][Chapters 3, 7]{https://doi.org/10.7907/z90z716m}.

We next turn to the question of pressure support.
Because the velocity scales as $r^{-1/2}$ and the density as $r^{-3/2}$, the ram pressure scales as $r^{-5/2}$.
By contrast, the gas pressure scales as $\rho T \propto r^{-17/8}$, which is a weaker scaling than the ram pressure.
As a result if the ram pressure dominates over the gas pressures at any radius it dominates everywhere inside that radius.
With equation~\eqref{eq:rbondi} and the relation
\begin{align}
	P = \gamma^{-1} \rho \cs^2
\end{align}
we find
\begin{align}
	P_{\mathrm{agn, gas}} \leq P_{\mathrm{agn}} = \frac{2 G \mstar \rho}{\gamma\rbondi},
\end{align}
where $\gamma$ is the adiabatic exponent.
Comparing this to the ram pressure using equation~\eqref{eq:vscale} we find that the gas pressure is at most of order the ram pressure at the Bondi radius, and the ram pressure rapidly comes to dominate further inwards.

The radiation pressure follows a similar relation.
With
\begin{align}
	P_{\mathrm{rad}} = \frac{1}{3} a T^4
\end{align}
and equation~\eqref{eq:tscale} we find
\begin{align}
	P_{\mathrm{rad}} \propto r^{-5/2},
\end{align}
which is the same scaling as the ram pressure.
Hence either the ram pressure dominates over radiation pressure at all radii or else the reverse holds.
We can therefore evaluate which one dominates at the stellar radius, giving
\begin{align}
	\frac{P_{\mathrm{rad}}}{P_{\mathrm{ram}}} = \frac{a \tstar^4\rstar }{3 \rho G \mstar}.
\end{align}
Using equation~\eqref{eq:tscale} we have
\begin{align}
\partial_r T^4 = -\frac{5 T^4}{2 r},
\end{align}
so we can write $T^4$ in terms of the luminosity and obtain
\begin{align}
	\frac{P_{\mathrm{rad}}}{P_{\mathrm{ram}}} \approx \frac{a \kappa L}{160 \pi \sigma G \mstar},
\end{align}
where $L$ here is the luminosity in the accretion stream.
With $a=4\sigma/c$ and equation~\eqref{eq:ledd} we find
\begin{align}
	\frac{P_{\mathrm{rad}}}{P_{\mathrm{ram}}} \approx  \frac{L}{10 \ledd}
\end{align}
Hence radiation pressure is comparable to the ram pressure only when $L$ is large compared with $\ledd$.
This does occur in our models, likely driving super-Eddington outflows, and so is cause for some caution and further study.
Nonetheless we suspect that the order of magnitude of the boundary conditions is not substantially altered by this.
This is partly because we expect aspherical effects where radiation is concentrated in some regions and attenuated in others, so it seems likely that some material is accreted at a velocity comparable to that in equation~\eqref{eq:vscale}, even if at other longitudes and latitudes material is flowing out at $\vesc$.

The remaining assumptions can be verified with order of magnitude estimates.
We have claimed that the luminosity within the stream is constant.
To verify this we begin with the first law of thermodynamics in the form
\begin{align}
	Tds = dE + PdV,
\end{align}
where $E$ is the total specific energy including kinetic and potential and $V$ is the specific volume.
With the ideal gas law we can write
\begin{align}
	PdV = -\frac{P}{\rho} d\ln \rho = -\frac{\cs^2}{\gamma} d\ln\rho.
\end{align}
Assuming the stream is not pressure supported, potential energy just turns into kinetic energy and
\begin{align}
	dE = c_\mathrm{p} dT \approx \cs^2 d\ln T.
\end{align}
Because $\rho$ varies more rapidly than $T$ we see that $|PdV| \gg |dE|$.
So
\begin{align}
	\frac{dL}{dm} = -T\frac{ds}{dt} \approx -P\frac{dV}{dt} = \frac{\cs^2}{\gamma} \frac{d\ln\rho}{dt}.
\end{align}
To integrate this through the stream we note that
\begin{align}
\int \cs^2 \frac{d\ln \rho}{dt} dm = \int \frac{dr}{dt} \cs^2 \frac{dm}{dr} \frac{d\ln \rho}{dm} dm = \int 4\pi r^2 \rho v \frac{-3 \cs^2}{2r} dr = \mdotbondir \int \frac{-3 \cs^2}{2r} dr  \propto \int r^{-1} T dr \propto \int r^{-13/8} dr.
\end{align}
With this we obtain
\begin{align}
	\Delta L \approx \mdotbondir \csa^2 \left(\frac{\rbondi}{\rstar}\right)^{5/8}.
\end{align}
This is less than the luminosity jump at the shock, which is given by equation~\eqref{eq:Lshock}
\begin{align}
	L_{\mathrm{shock}} \approx \mdotbondir v^2 \approx \dot{M} \csa^2 \frac{\rbondi}{\rstar}.
\end{align}
As the luminosity in the stream is given by the intrinsic luminosity of the star plus that of the shock, so long as $\lstar > 0$ the luminosity before the shock changes by at most order unity.

To show that the mass of the stream is small compared with the mass of the star, note that the mass of the stream is given by
\begin{align}
	M_{\mathrm{stream}} = \mdotbondir t_{\mathrm{ff}} < \mdotbondi t_{\mathrm{ff}},
\end{align}
where $t_{\mathrm{ff}}$ is the free-fall time from $\rbondi$ to $\rstar$.
This time is of order $\sqrt{\rbondi^{3}/G \mstar}$, so using equation~\eqref{eq:mbondi} we find
\begin{align}
	M_{\mathrm{stream}} < \sqrt{\frac{16 \pi^2 \rbondi^7 \rhoagn^2 \csa^2}{G \mstar}} \approx \rbondi^3 \rhoagn.
\end{align}
Even with generous values of $\rbondi \approx 10^{15}\mathrm{cm}$ and $\rhoagn \approx 10^{-15}\mathrm{g\,cm^{-3}}$, $M_{\mathrm{stream}} < 10^{-3} \Msun$ and so is negligible compared with the mass of the star.

\begin{figure}
    \centering
    \includegraphics[width=0.5\columnwidth]{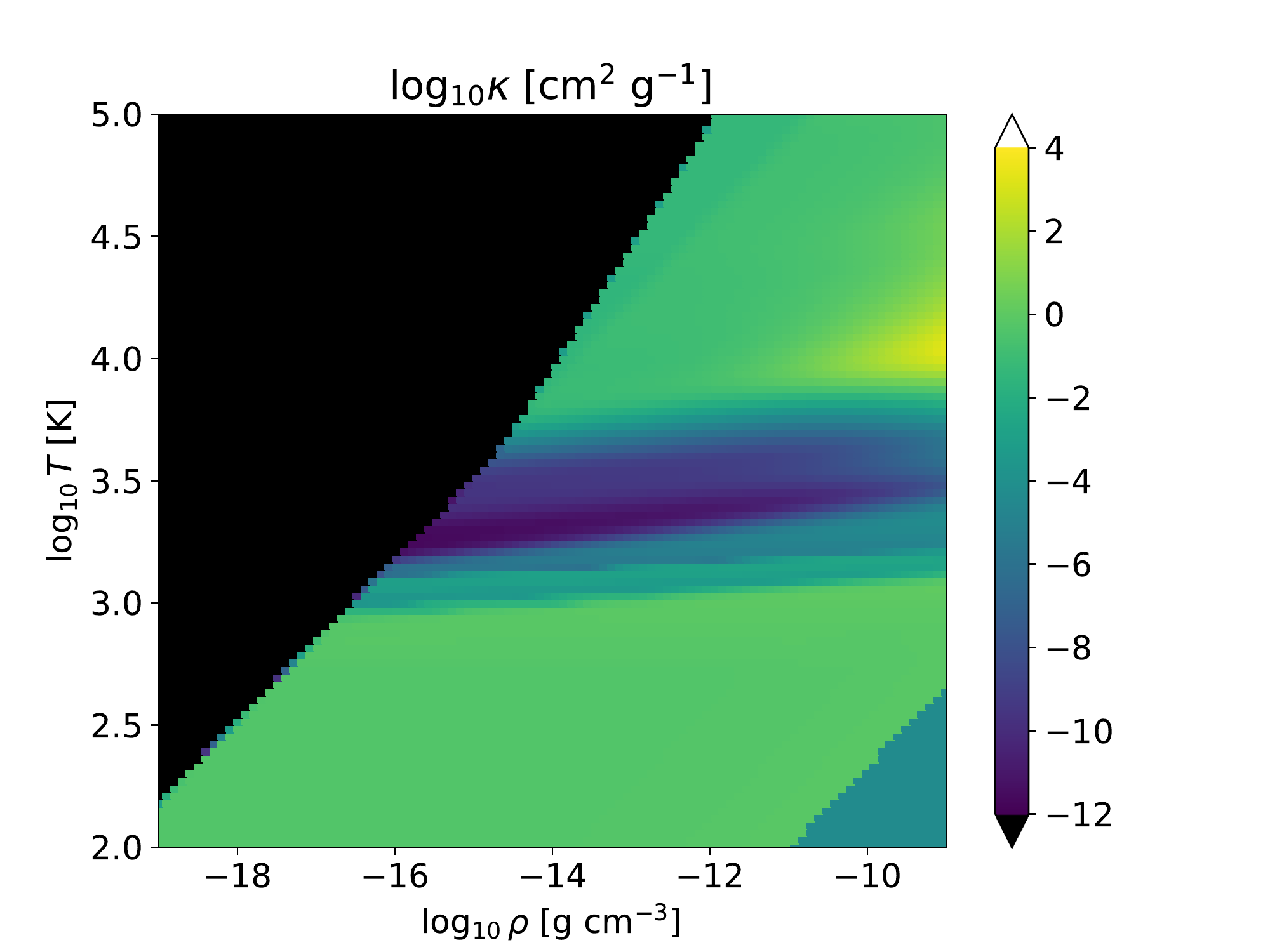} 
    \caption{The opacity for a mixture of $76\%$ Hydrogen, $22\%$ Helium, and $2\%$ solar-distributed metals is shown as a function of temperature and density.}
    \label{fig:kappa}
\end{figure}

We have also assumed that the opacity of the accretion stream is a constant in space, though it may evolve in time.
This is not true (see Figure~\ref{fig:kappa}), but for our purposes it suffices to be able to define an average value in the stream and have this not vary too strongly as the star evolves.
Because we are concerned with an average over the stream this may not be so bad an approximation, as the averaging process smooths over strong bumps and ridges in the opacity function.
On the other hand stellar evolution produces systematic trends in the temperature and density structure of the accretion stream, which in turn could translate into systematic trends in the mean opacity. As such this assumption should be treated with significant caution.

\section{Model Grids with different AGN sound speeds}
\label{sec:extragridcs}
\begin{figure*}[h!]
  \centering
  \subfloat{\includegraphics[width=0.5\columnwidth]{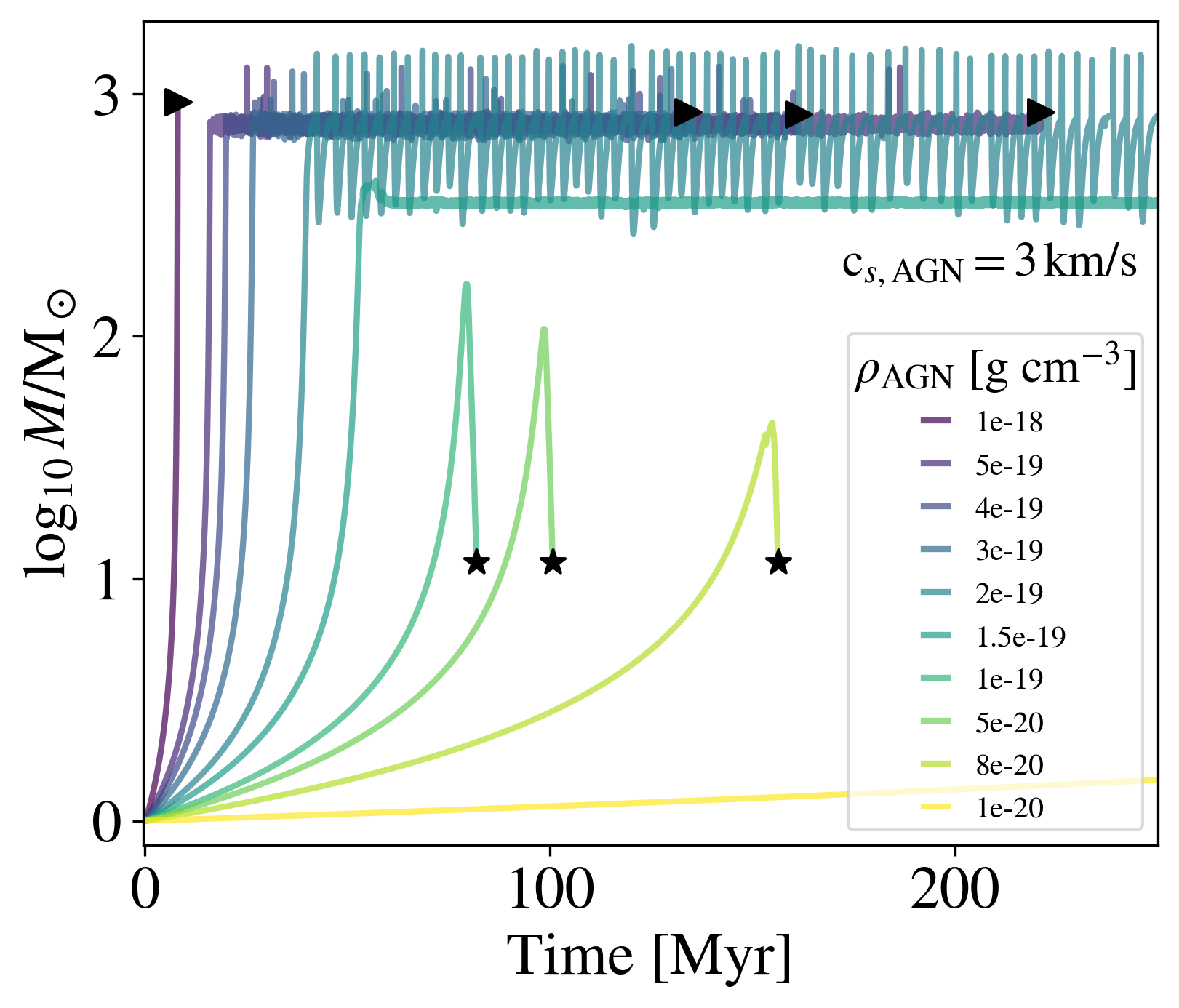}}\hfill
  \subfloat{\includegraphics[width=0.5\columnwidth]{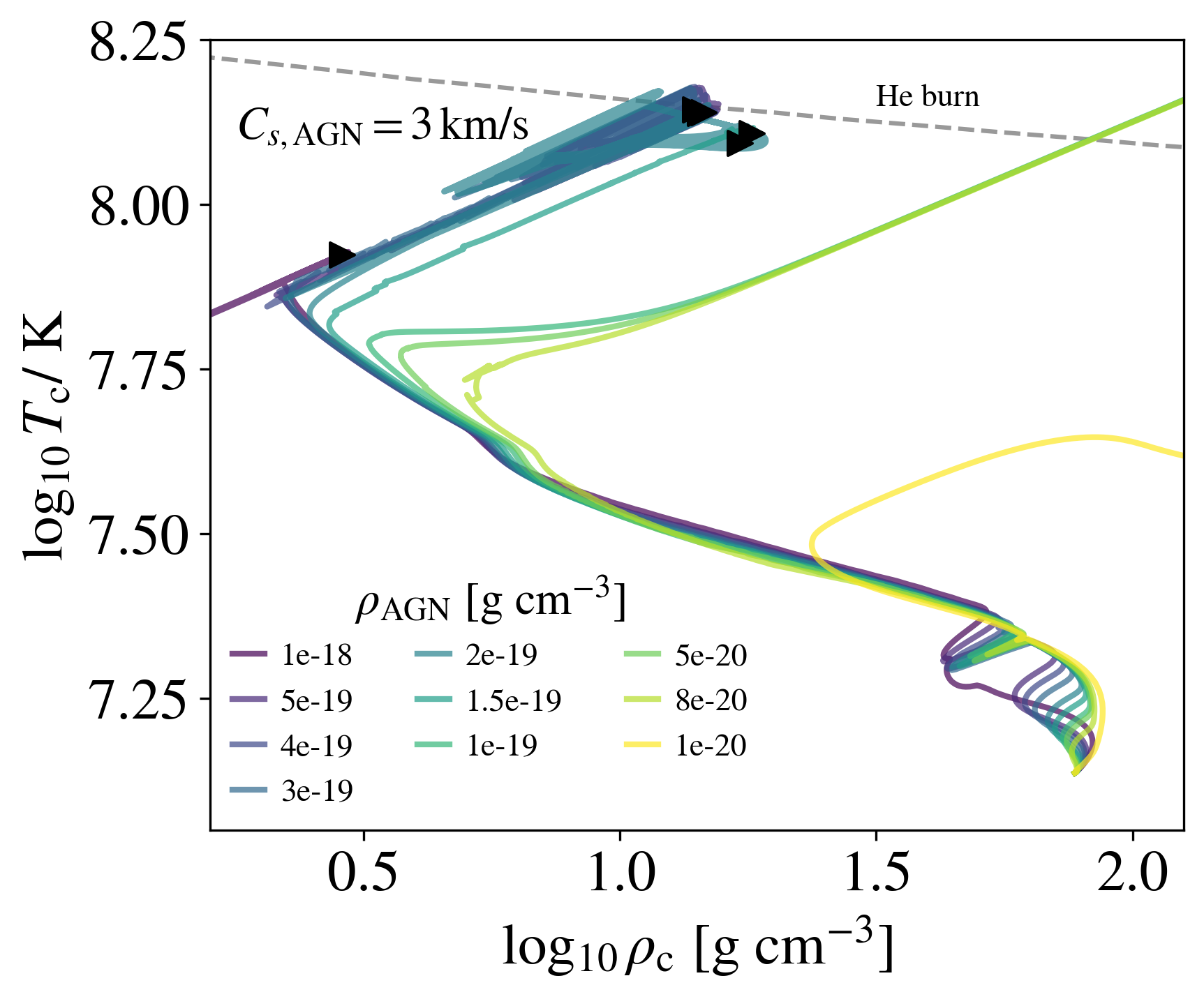}}
   \caption{A grid of stellar models evolved with a fixed AGN sound speed of $3 \, \kms$ and AGN densities ranging from $10^{-18}$ to $10^{-20}\,\mathrm{g\,cm^{-3}}$. The left panel shows the evolution of stellar mass as a function of time. Models evolving at densities higher than $\approx 10^{-19}\,\mathrm{g\,cm^{-3}}$ experience runaway accretion and become supermassive stars (tracks ending with triangle symbols). Models evolving at densities $10^{-19} \le \rhoagn \le 8\times 10^{-20}$ become massive stars before losing mass via super-Eddington winds and ending their lives with $M \approx 10\mso$ (tracks ending with a star symbol). At densities lower than $\approx 8 \times 10^{-20}\,\mathrm{g\,cm^{-3}}$ stars end their main sequence evolution before accreting sufficient material to become massive stars ($M \lesssim 8\mso$, tracks ending with a circle). The right panel shows a zoom of a central density -- central temperature plot. Ending symbols are the same in the two panels.}
   \label{fig:appendix_gridcs3}
\end{figure*}

\begin{figure*}[h!]
  \centering
  \subfloat{\includegraphics[width=0.5\columnwidth]{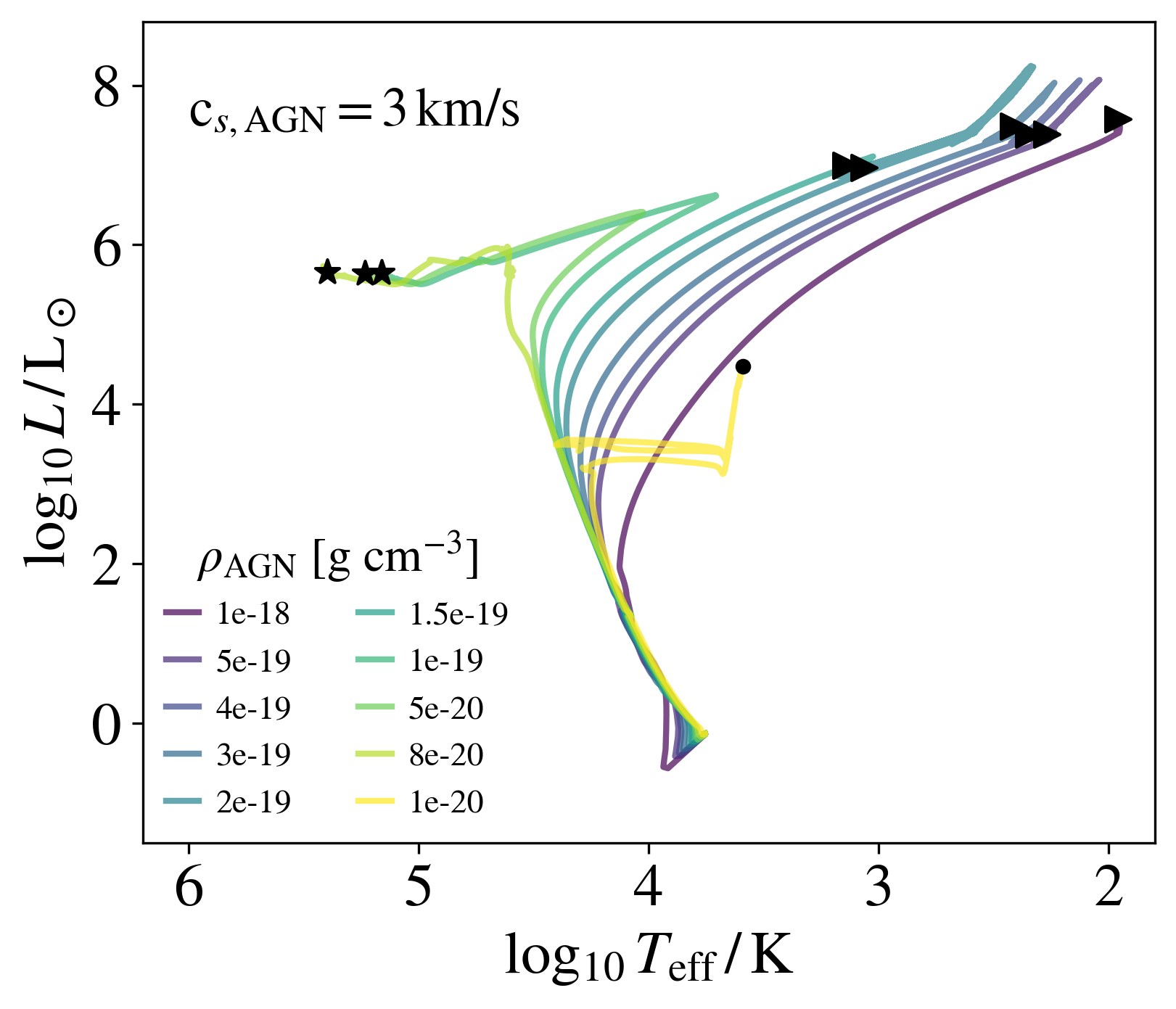}}\hfill
  \subfloat{\includegraphics[width=0.5\columnwidth]{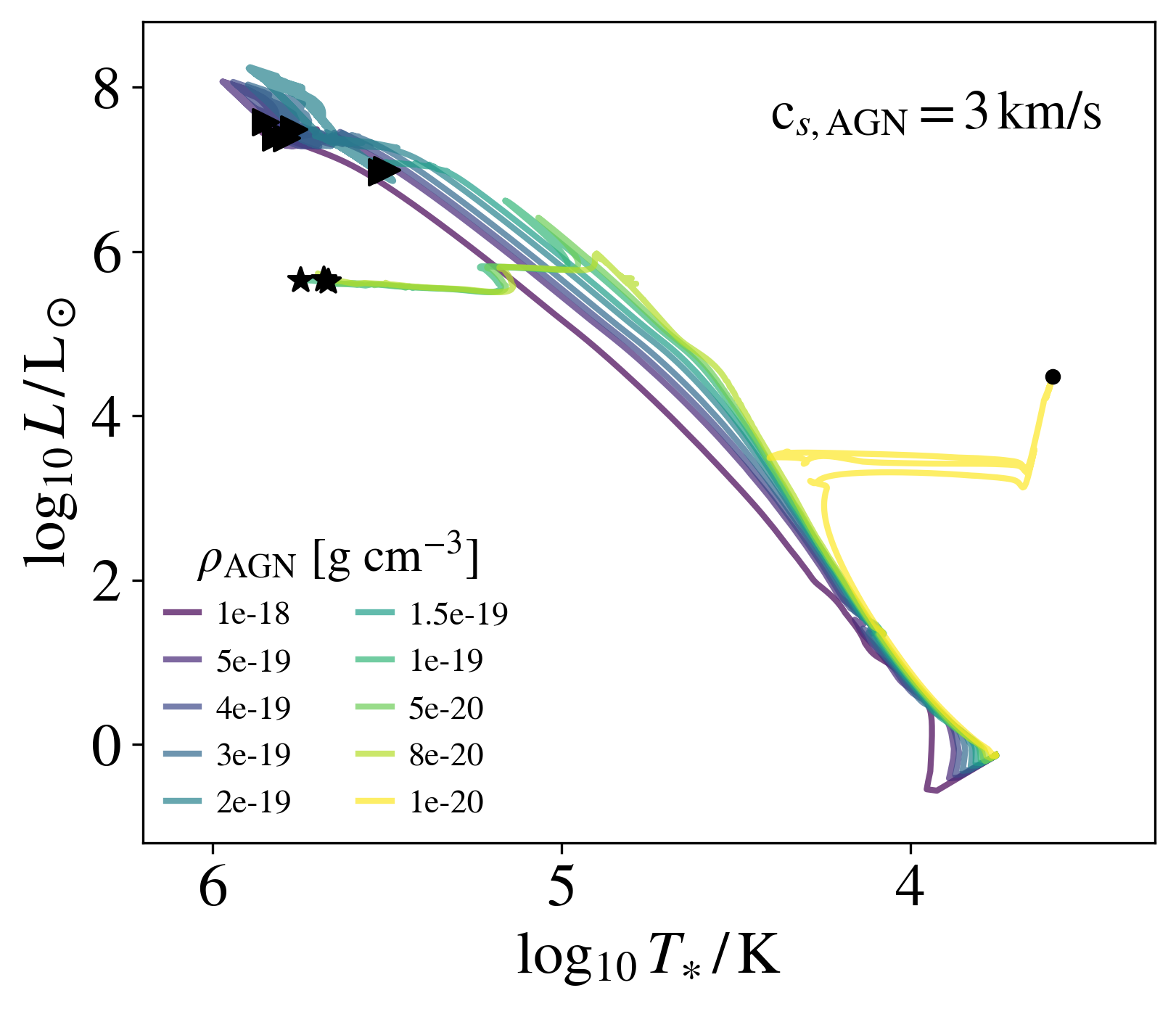}}
   \caption{Evolution on the HR-Diagram for the models shown in Fig.~\ref{fig:appendix_gridcs3}.
   Left panel: The effective temperature of the models is calculated assuming electron scattering opacity for the accretion stream. This is the temperature of the star for an observer sitting at the Bondi radius. Right panel: Same as left panel, but using the surface temperature of the MESA model instead of the effective temperature (Eq.~\ref{eq:Teff}).  This diagram is useful to understand the type of evolution \agns\ models are undergoing, as compared to  canonical stellar evolution. Note that the ending symbols are the same in the two panels.}
    \label{fig:appendix_gridcs3_hrd}%
\end{figure*}

\begin{figure*}[h!]
  \centering
  \subfloat{\includegraphics[width=0.5\columnwidth]{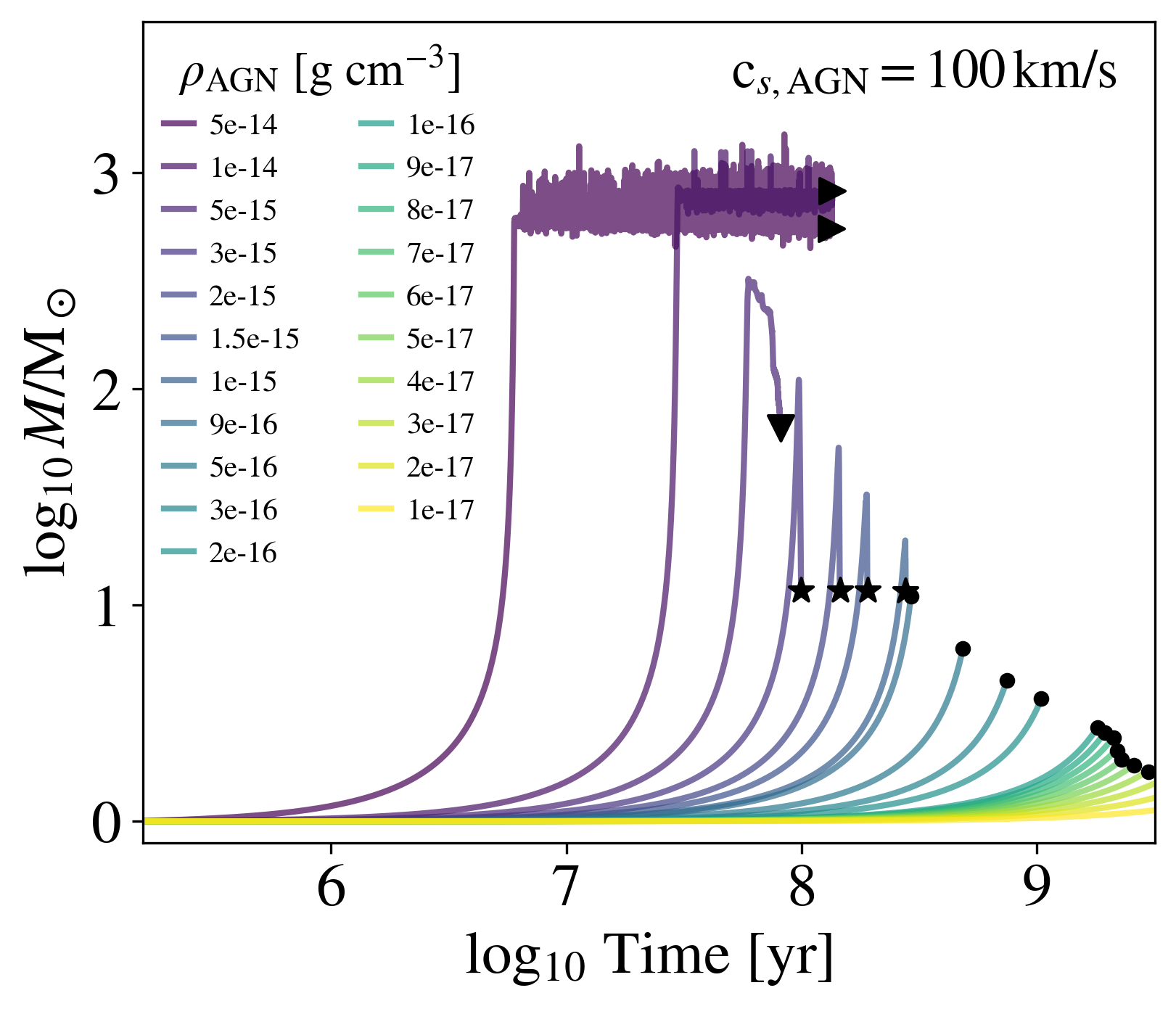}}\hfill
  \subfloat{\includegraphics[width=0.5\columnwidth]{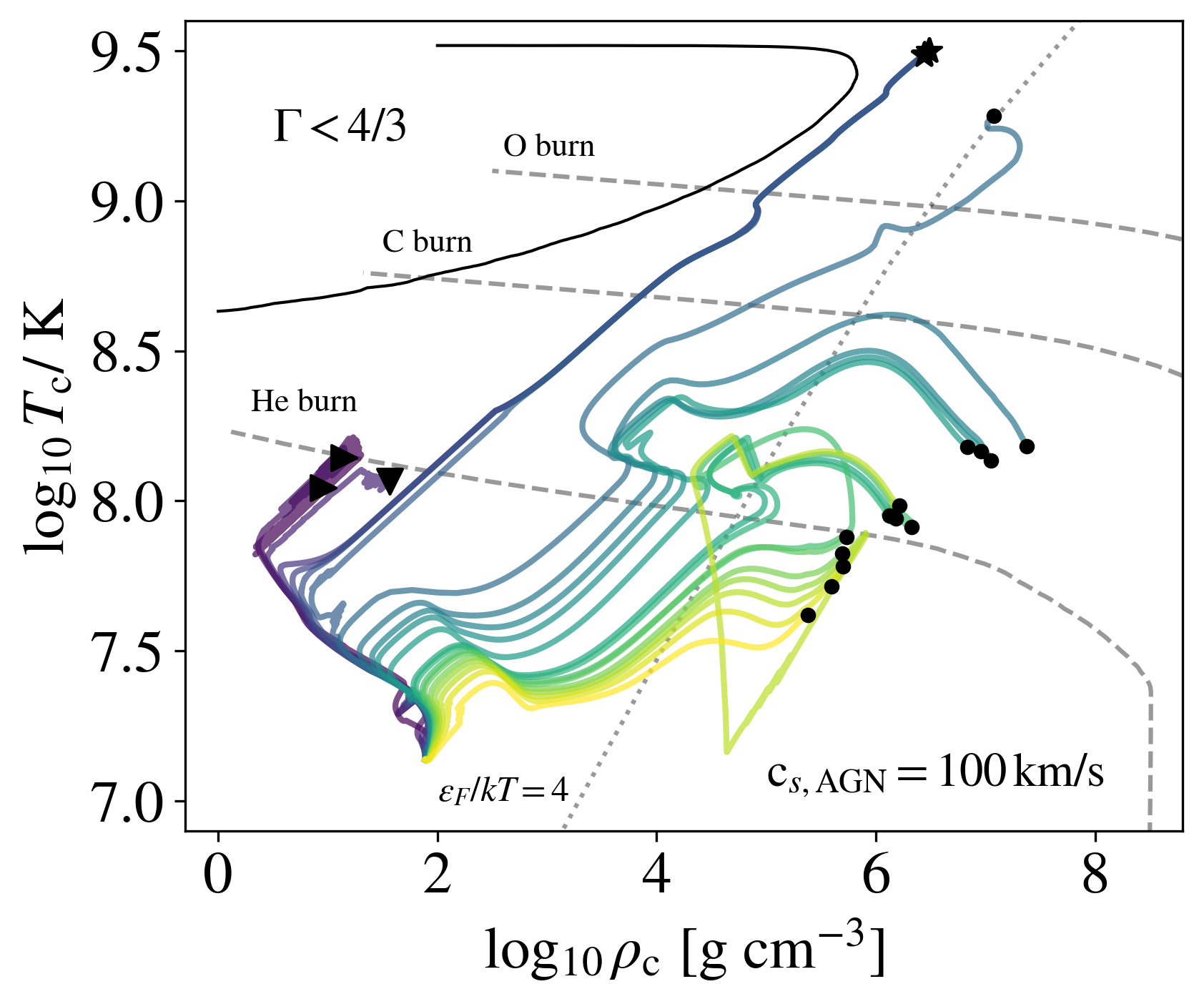}}
   \caption{A grid of stellar models evolved with a fixed AGN sound speed of $100 \, \kms$ and AGN densities ranging from $5\times10^{-14}$ to $10^{-17}\,\mathrm{g\,cm^{-3}}$. The left panel shows the evolution of stellar mass as a function of time. Models evolving at densities higher than $\approx 5\times10^{-15}\,\mathrm{g\,cm^{-3}}$ experience runaway accretion and become supermassive stars (tracks ending with triangle symbols). Models evolving at densities $5 \times 10^{-15} \le \rhoagn \le 5\times 10^{-16}$ become massive stars before losing mass via super-Eddington winds and ending their lives with $M \approx 10\mso$ (tracks ending with a star symbol). At densities lower than $\approx 5 \times 10^{-16}\,\mathrm{g\,cm^{-3}}$ stars end their main sequence evolution before accreting sufficient material to become massive stars ($M \lesssim 8\mso$, tracks ending with a circle). The right panel shows the evolution on a central density -- central temperature plot. Ending symbols are the same in the two panels.}
    \label{fig:appendix_gridcs100}%
\end{figure*}

\begin{figure*}[h!]
  \centering
  \subfloat{\includegraphics[width=0.5\columnwidth]{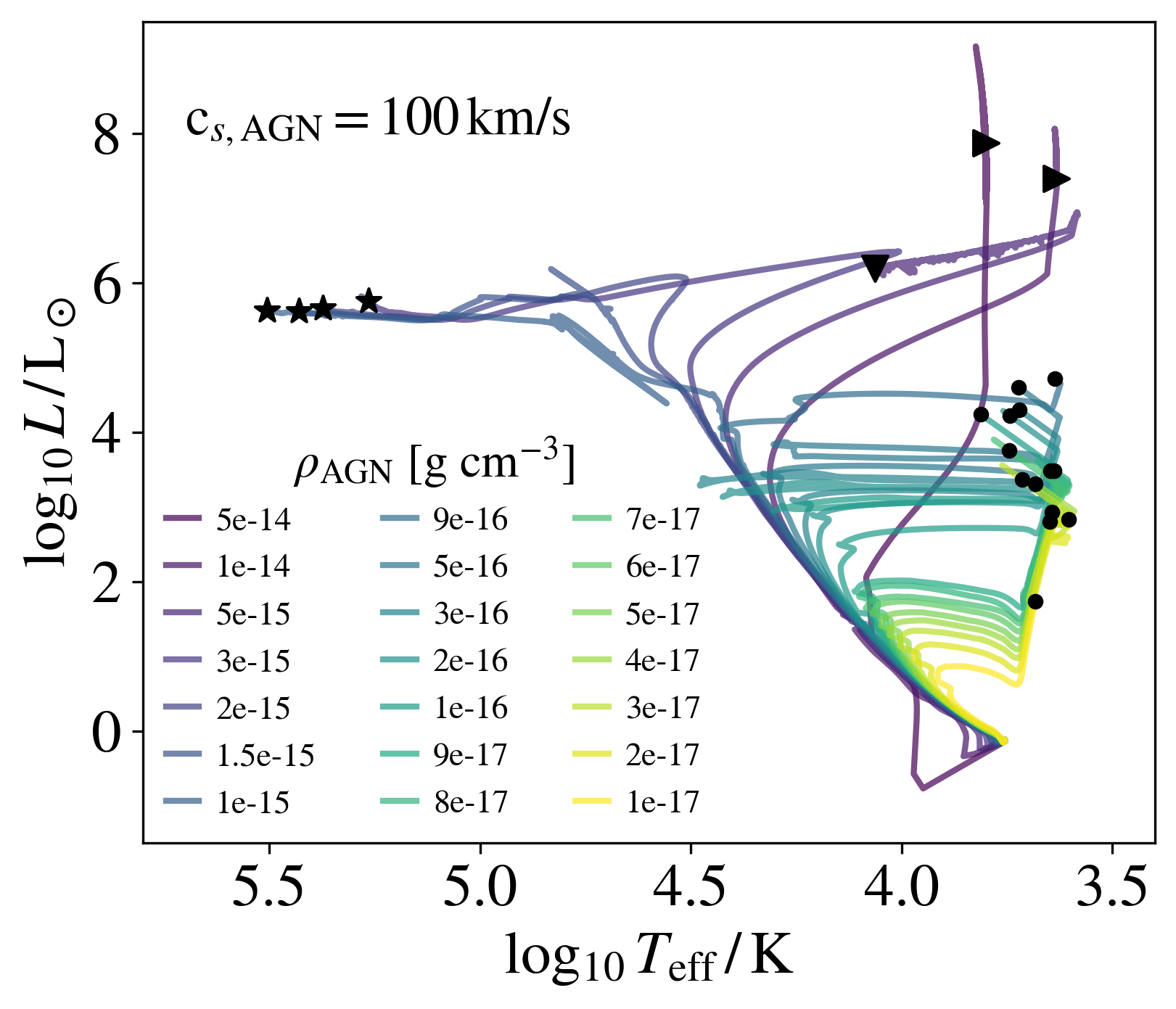}}\hfill
  \subfloat{\includegraphics[width=0.5\columnwidth]{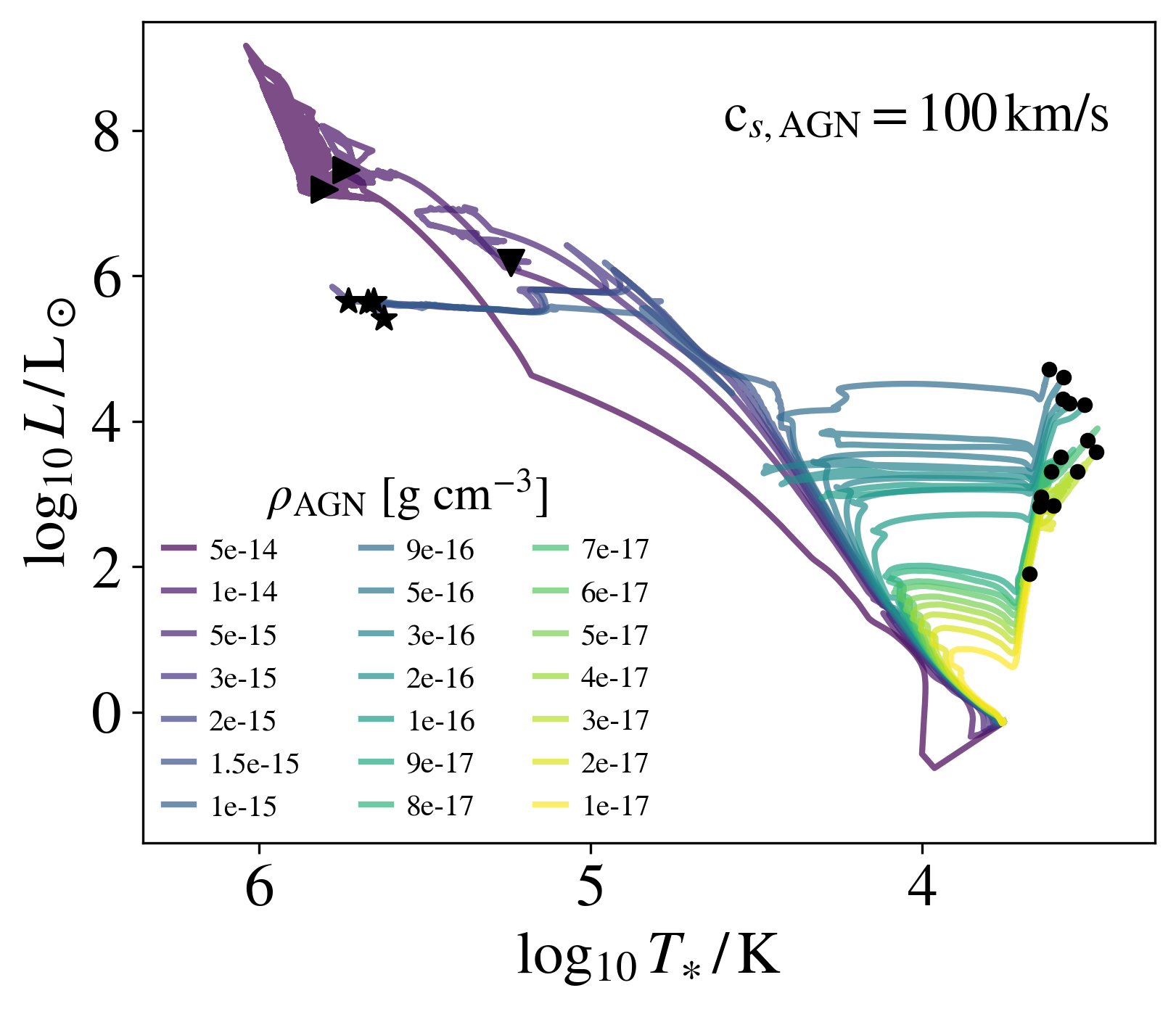}}
   \caption{Same as Fig.~\ref{fig:appendix_gridcs3_hrd} but for the models shown in Fig.~\ref{fig:appendix_gridcs100}.}
    \label{fig:appendix_gridcs100_hrd}%
\end{figure*}

Here we show model grids for two different assumptions of the AGN sound speed, $\csa = 3 \, \kms$ (Fig.~\ref{fig:appendix_gridcs3} \& \ref{fig:appendix_gridcs3_hrd}) and $\csa = 100 \, \kms$ (Fig.~\ref{fig:appendix_gridcs100} \& \ref{fig:appendix_gridcs100_hrd}). 

\bibliographystyle{aasjournal}
\bibliography{biblio.bib}

\end{document}